\RequirePackage{fix-cm}
\documentclass[smallextended]{svjour3}       
\smartqed  
\usepackage{graphicx}
\usepackage{algpseudocode}
\usepackage{algorithmicx}
\usepackage{algorithm}
\usepackage{subcaption}
\usepackage{caption}
\begin{document}

\title{Friendship and Selfishness Forwarding: applying machine learning techniques to Opportunistic Networks data forwarding
}


\author{Camilo Souza         \and
        Edjair Mota         \and
		Leandro Galvao      \and
		Diogo Soares		\and
        Pietro Manzoni      \and
        Juan Carlos Cano    \and
        Carlos Calafate     \and 
}


\institute{Camilo Souza, Edjair Mota, Leandro Galvao, Diogo Soares \at
Institute of Computing, Federal University of Amazonas -- Manaus -- AM -- Brazil-BR\\
              \email\{camilo.souza,edjair,galvao,diogo.soares\}@icomp.ufam.edu.br
           \and
           Pietro Manzoni, Juan Carlos, Carlos Calafate \at
Computer Engineering Department, Universitat Politecnica de Valencia -- Valencia -- Spain -- 46022\\
           \email\{pmanzoni,jucano,calafate\}@disca.upv.es
}

\date{Received: date / Accepted: date}

\maketitle

\begin{abstract}
Opportunistic networks could become the solution to provide communication support in both cities where the cellular network could be overloaded, and in scenarios where a fixed infrastructure is not available, like in remote and developing regions. A critical issue that still requires a satisfactory solution is the design of an efficient data delivery solution. Social characteristics are recently being considered as a promising alternative. Most opportunistic network applications rely on the different mobile devices carried by users, and whose behavior affects the use of the device itself.

This work presents the ``Friendship and Selfishness Forwarding'' (FSF) algorithm. FSF analyses two aspects to make message forwarding decisions when a contact opportunity arises: First, it classifies the friendship strength among a pair of nodes by using a machine learning  algorithm to quantify the friendship strength among pairs of nodes in the network. Next, FSF assesses the relay node selfishness to consider those cases in which, despite a strong friendship with the destination, the relay node may not accept to receive the message because it is behaving selfishly, or because its device has resource constraints in that moment. 

By using trace-driven simulations through the ONE simulator, we show that the FSF algorithm outperforms previously proposed schemes in terms of delivery rate, average cost, and efficiency. 
\end{abstract}

\section{Introduction}
\label{intro}

In the last five years, activities related with Opportunistic Networks (OppNets) \cite{tang:2012} have largely grown, attracting the attention of the scientific community \cite{7823357}. This network paradigm exploits any
connectivity that arises through the movement of the nodes to distribute content, being largely based on the store-carry-forward mechanism derived from the Delay/Disruption Tolerant Networking (DTN) paradigm \cite{li:2009b}. 
Efficient data distribution is a challenging research topic in Opportunistic Networks
due to the presence of long delays and frequent disconnections caused by link failures, network partitioning, topology changes, or node mobility \cite{carina:2007}.

In the literature and  derived from the DTN world, we find several distribution algorithms for OppNets. Most of these  algorithms belong to three different categories \cite{zhu2013survey}: (i) message-ferry-based, (ii) purely opportunistic based, and (iii) prediction-based. Recently, researchers addressed this issue from a social perspective \cite{xia2015socially}, and a new category for data distribution protocols emerged. In  most of these works, different metrics are introduced to take into account social characteristics like friendship, altruism, selfishness, popularity, and centrality.

In this paper, we are interested in DTN-based message distribution by taking into account two social characteristics: (i) the friendship level between the nodes, and (ii) the individual selfishness of the message relay candidate. However, differently from  most of the works in the literature, we have modeled these two social aspects as a classification problem, applying machine learning techniques to these tasks. Based on these assumptions, we propose the ``Friendship and Selfishness  Forwarding'' (FSF) algorithm, which provides a novel approach based on machine learning for classifying the friendship strength among two nodes, as well as the selfishness of the message relay candidate. 

To validate our proposal, we develop a set of experiments that include simulating the message delivery process in two scenarios. We then compare the obtained results in terms of delivery rate, average delay, cost, and efficiency against other protocols found in the literature, such as BUBBLE-RAP \cite{hui:2011}, Friendship Routing (FR) \cite{bulut2010friendship},
Socially Selfish Aware Routing (SSAR) \cite{li:2010}, Probabilistic Routing Protocol using History of Encounters and Transitivity (PROPhET) \cite{lindgren:2003}, and Epidemic \cite{vahdat:2000}.

The main contributions of this paper are:

\begin{itemize}
	\item Formalization and classification of the concepts of ``friendship'' and ``selfishness'' using machine learning techniques. As mentioned above, most of the works that take the friendship and selfishness among nodes into account introduce new metrics based on some features related to these characteristics. In this work we have used machine learning techniques, highlighting their efficiency in dealing with these social behaviors.
	
	\item Classification of the selfish behavior. We assume two types of selfish user behavior: those nodes that are only selfish in some situations (e.g., due to resource constraints), and those nodes that, by default, behave selfishly all the time. Most selfishness-based routing protocols do not take these two types of user selfishness into account, while FSF is able to deal with these two different behaviors concurrently.
	
	\item Dynamic classification of the friendship strength and selfishness type. FSF tries to adapt itself to the real conditions of each node during a contact. The Naïve Bayes machine learning algorithm is the core of this search towards a more realistic routing protocol.
	
\end{itemize}

The rest of this paper is organized as follows: in Section \ref{sec:related_works} we discuss some related works. Section \ref{sec:FSR_algorithm} details the proposed FSF protocol for OppNets applications. In Section \ref{sec:evaluation}, we describe the evaluation methodology. In Section \ref{sec:results}, we present and discuss the experimental results. Finally, in Section \ref{sec:conclusions}, we conclude the paper and refer to future works.

\section{Related Works}
\label{sec:related_works} 
Several forwarding algorithms are available in the OppNets literature. These algorithms
can be classified according to the number of copies of a single message
in the network \cite{spyropoulos2008efficient}. If, at any given time, there is only
one node in network that carries a copy of the message, the forwarding algorithm is
classified as single-copy algorithm; on the contrary, it is considered a multicopy algorithm. 
Zhu et al. \cite{zhu2013survey} classify the forwarding algorithms according to its 
approaches in three categories: (i) message-ferry-based, (ii) opportunity-based, and (iii) 
prediction-based. In the last years, protocols based on social relationships emerged as a fourth category. 

Message-ferry-based routing algorithms carry out data distribution in a network by
combining the store-and-forward mechanism with the use of additional nodes, called 
ferries, that act as data mules \cite{zhu2013survey}. Many routing algorithms have embraced this forwarding paradigm \cite{shah2003data,zhao2004message,zhao2005controlling,zhao2003message}, including
Burns et al. \cite{burns2008mora} who proposes a routing algorithm based on the observed meetings between peers, and the visits of peers to geographic locations. Message-ferry-based routing algorithms share a common problem: the overhead and extra cost to control the ferries.

In opportunistic-based algorithms the nodes exchange data with each other when a contact 
occurs, i.e., whenever they are within communications range. The Epidemic routing 
algorithm, proposed in \cite{vahdat:2000}, is an example of an algorithm based on 
this forwarding paradigm. This routing algorithm assumes that, when a pair of nodes 
is in the same coverage area, they send all the messages stored in their buffers to each other, increasing the message delivery probability.

Based on an opportunistic-based paradigm, some researchers have explored the network behavior to propose forwarding algorithms. For instance, Shaghaghian and Coates \cite{shaghaghian2015optimal} have used some simplifying assumptions about the network behavior to propose two forwarding algorithms. The main goal is to minimize the expected latencies from any node in the network to a particular destination in some situations. The authors take into account that forwarding algorithms for opportunistic networks should result in low-average latency and an efficient usage of network resources. Based on simulation results, the authors confirm  that their proposed algorithms are able to improve both the latency and the delivery rate.

In the same way, inspired by the Nash bargaining solution, Li et. Al proposes in
\cite{li2016novel} the GameR forwarding algorithm for OppNets. This algorithm 
uses a utility function derived from the estimated resource utilization ratio and 
the history delivery predictability. According to their obtained results, GameR can
efficiently utilize the network resources, improving the delivery ratio and 
decreasing the overhead on nodes under resource-constrained situations. Other proposals
based on this paradigm include \cite{spyropoulos:2005,juang2002energy,burgess:2006} and \cite{spyropoulos2008efficient}.

Prediction-based algorithms are a refinement of the opportunistic-based approach. These algorithms select the next-hop node based on some metrics. Generally, these metrics
make use of characteristics such as probability of message delivery to the destination node \cite{lindgren:2003,liu2009optimal,nguyen2007probabilistic}, or based on the encounters history of the receiving node \cite{burgess:2006,oliveira:2009,zhang2014novel}.

The routing protocol proposed in this work belongs to the social-based category. Several researchers have regarded the use of some social characteristics in the design
of new DTN routing protocols because most DTN applications include a huge number of mobile devices carried by humans, and, according to Zhu et al. \cite{zhu2013survey},
the behavior representing the utilization of devices is better described by social network models. Among the social characteristics worth considering we have altruism \cite{okasha:2005}, friendship \cite{bulut:2010,qin2015nfcu,souza2016fsf}, selfishness \cite{li:2010,souza2016fsf}, centrality, and popularity \cite{hui:2011}.

In \cite{bulut2010friendship}, Bulut et al. proposed a routing algorithm based on friendship. To model the friendship among nodes, the authors considered the frequency of 
contacts, their longevity, and their regularity. The authors claim that a pair of nodes are friends if they regularly have long and frequent contacts. Based on these assumptions,
a new metric called Social Pressure Metric (SPM) is proposed to represent the social pressure motivating friends to visit each other and share their experiences. Different 
from \cite{bulut2010friendship},  Qin et al. proposed in \cite{qin2015nfcu} the New Friendship-based routing with buffer management based on the Copy Utility (NFCU) algorithm. 
This algorithm takes into account improvements in the friendship-based routing including in the selected metric an analysis of the inherent drawbacks such as the consideration of the contact periods when constructing the SPM metric, and the buffer management issue.

In our work, similarly to the works mentioned above, we  model friendship by taking into account the average contact duration and frequency; in addition, we consider the 
number of calls and the text messages exchanged since we believe such information is a clear evidence of friendship among two nodes. In the real world, it is reasonable to 
suppose that the more calls and/or messages exchanged by a pair of nodes, the stronger will be the friendship among them. Moreover, our algorithm uses a machine learning 
approach derived from real-world examples.

Our FSF routing algorithm also takes into account the selfishness of the message 
relay candidate in forwarding decisions. According to Miao et al. 
\cite{miao2013investigation}, most existing routing protocols take for granted that
nodes take part in the routing process. However, people in the real world are socially selfish, and only tend to collaborate with a restricted set of people. Based on 
these assumptions, Li et al. have proposed a routing algorithm based on node 
selfishness \cite{li:2010}. The Social Selfishness Aware Routing (SSAR) is a 
protocol that introduced selfishness considerations in DTN scenarios. This routing 
algorithm tries to compensate for performance loss by allocating resources (buffers and bandwidth) based on packet priority. The authors of that paper consider that selfishness issues should be integrated into new routing algorithms since people are socially selfish, that is, they are willing to forward messages to a limited number of persons, and this willingness depends on the strength of the relationship among them. For example, if the social bonding between node A and node B is strong, node A will always accept messages to node B. The routing algorithm proposed in our work considers that, in several situations, node A may not accept those messages whose destination node is node B for the sake of saving resources, no matter how strong is their mutual friendship. Besides that, in this work we have also considered two types of user selfish behavior: those nodes that are selfish only in some situations (e.g. resources constraints), and those nodes behaving selfishly all the time. This is an important issue because some nodes are selfish only in some situations, but most of the selfish-based routing protocols assume that nodes will be selfish in all situations.

Below, we detail the fundamental characteristics of our FSF algorithm.

\section{The FSF algorithm}
\label{sec:FSR_algorithm} 

In this section, we introduce the FSF forwarding strategy by providing all the details necessary to fully understand its forwarding strategy and characteristics.

\subsection{FSF Forwarding Strategy}\label{subsec:forwarding_strategy}

Algorithm \ref{alg1} shows the pseudocode description of how the FSF forwarding strategy
works. The FSF strategy seeks to perform the message forwarding process from two points
of view: the source side and the relay side. From the source side point of view, FSF
searches the best routes for a message to its destination. To make this decision, 
FSF takes into account the friendship strength among nodes. Our FSF technique assumes  
that, the greater the friendship strength, the bigger will be the message
delivery probability. From the relay side point of view, FSF seeks to consider the 
cases in which the relay node rejects receiving the message (because he/she
is selfish, or his/her device has resource constraints in that moment). In this way,
and to meet these assumptions, FSF performs two steps when a contact opportunity arises:

\begin{algorithm}[h]
	\caption{\textit{Friendship and Selfishness Forwarding Algorithm}}\label{alg1}
	\begin{algorithmic}[1]
		\Procedure{FSF}{$mDest,cToRelay$}
		\State $friendshipStrength \gets NaiveBayesClassifier(mDest,cToRelay)$
		\If{$friendship,strength  ==   $  "strong"}
		
		$ SelfishnessAssessment(cToRelay)$
		\Else
		
		$ Message Will Not Be Forwarded$
		\EndIf
		\EndProcedure
	\end{algorithmic}
\end{algorithm}

\begin{itemize}
	\item step (i) classification of friendship strength among the nodes -- To do it, FSF invokes the NaiveBayesClassifier() method. 
	If this method assesses the friendship strength as ``weak'', the message will not 
	be forwarded and FSF will stop. On the other hand, if the friendship strength is 
	classified as ``strong'', FSF invokes Algorithm \ref{alg3} to run step (ii) to 
	assess the relayer's selfishness.
	
	\item step (ii) assessment of relayer's selfishness -- It is achieved through a reputation
	system. We establish the node reputation as either selfish or not selfish. FSF also
	takes into account that selfish nodes can still behave in two manners: individually selfish or socially selfish. If the node is assessed as individually selfish, the message 
	will never be forwarded. On the other hand, socially selfish nodes will relay
	messages if, and only if, the message destination belongs to the same community as the
	message relay.
	
	It is worth mentioning that FSF also takes into account that no selfish nodes can behave as individually selfish in cases in which his/her devices have resource 
	constraints. In this work, the device resources analyzed by FSF are: (1) the battery
	level (represented by $\alpha$), and (2) the memory occupation (represented by $\beta$).
	Thus, from a node resource constraints perspective, a non-selfish node will receive 
	a message if, and only if, his/her device resources have a battery level above 
	$\alpha$, and/or if memory occupation is less than $\beta$. More details about this issue
	will be provided in subsection \ref{not:selfish}.
	
\end{itemize}

Below, we further detail steps (i) and (ii) performed by FSF in the message forwarding process.

\subsection{Step (i): classification of the friendship strength}\label{subsec:classification}

The first step of the FSF algorithm is to classify the friendship strength among the message destination to the candidate relay. Differently from other works that take friendship strength into account when taking routing decisions \cite{bulut:2010,qin2015nfcu}, we have applied a machine learning technique to classify the friendship strength among the nodes, called Naive Bayes classifier \cite{rish2001empirical}.

We choose a Naive Bayes classifier based on two facts: first, the promising performance reported in other works found in the literature like \cite{pang2002thumbs,mccallum1998comparison,lewis1998naive}. Second, the nature of the problem itself, for which it is reasonable to assume that the attributes are statistically independent. A problem is statistically independent when a value for one attribute does not imply another value for another attribute. For example, if in a problem we have two attributes, i.e., amount of calls and amount of text messages, the statistical independence of these attributes means that, if two nodes called each other several times, it does not imply that several text messages were also exchanged. According to \cite{gama2012aprendizagem,lewis1998naive,panda2007network}, the Naive Bayes classifier achieves excellent results in situations of statistical independence of the attributes used for the classification task, that is, it can correctly predict the class of a given instance.

Before moving ahead, we shall give more details about some important concepts required to understand how FSF performs its first task. Basically we will describe the concept behind a \textit{database}, \textit{attributes}, and \textit{classes} from a machine learning perspective. Next, we explain how the Naive Bayes classifier works, and finally we present an example of a Naive Bayes classification.

\subsubsection{Training and testing data for the Naive Bayes classifier}\label{subsubsec:training_naive_bayes}

To apply a machine learning technique, it is necessary to have a ``database'' for training and testing, including all the ``attributes'' considered significant for the problem itself \cite{amorim2008tecnicas} and its respective ``class''. From a machine learning perspective, the \textbf{class} is the concept you want to learn, and the \textbf{attributes} are the ``evidences'' that can help at predicting the class. 

For example, to classify the friendship strength among two nodes in a network we will consider as evidences to be used in this problem, and their possible values, the number of contacts (\textit{weak}, \textit{average}, \textit{higher}), the contacts' duration (\textit{small}, \textit{average}, \textit{large}) and the contact frequency (\textit{weak}, \textit{higher}).

Also, we will consider that the possible classes of the problem are: \textbf{weak} friendship (i.e., they are not friends) and \textbf{strong} friendship (i.e., they are friends). 

The \textbf{database} for this problem can therefore be represented by several tuples containing the values relative to each 
attribute used, and the corresponding class according to these attributes. In our case, the training and testing data will be a database consisting of tuples including a value for the number of contacts, contacts duration and contact frequency, and a value for the corresponding class.

\begin{figure}
	\includegraphics[width=0.75\textwidth]{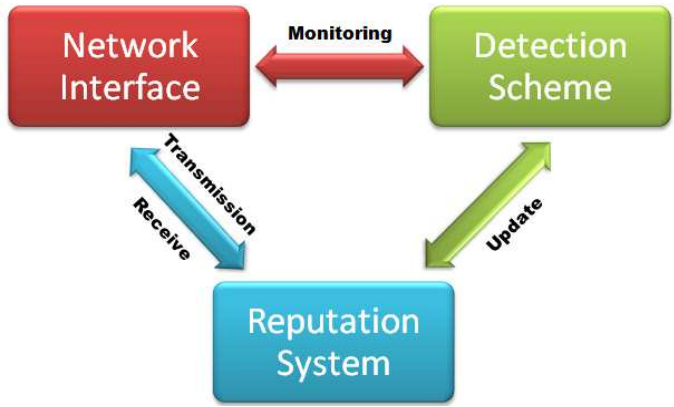}
	\caption{The selfish classification system running at each network node.}
	\label{fig:1}       
\end{figure}

In this work the training and testing data used for the friendship strength classification were derived from an MIT researcher experiment \cite{pentland2009inferring}. That 
project got the data from a test with mobile devices equipped with a Bluetooth network interface. The experiment collected data from 97 mobile phones over the course of six 
months. We can also have access to information related to connectivity, proximity, location, and activities of these users.

The main motivation for using this trace is the call history and text message record among the users who took part in the experiments. These informations, together with the frequency of meetings and contact duration among nodes, are the basis for classifying the strength of the friendship among each pair of nodes in the network. 

Each tuple belonging to the database used in this work as training and testing data is divided into five attributes. These attributes are defined as follows:

\begin{itemize}
	\item column 1: frequency of meetings (FM) -- represents how many times two nodes were within the radio range of each other.
	
	\item column 2: contact duration (CD) -- represents the average time that two nodes remain within each other's coverage area.
	
	\item column 3: amount of calls (AC) -- represents the number of calls exchanged by a pair of nodes.
	
	\item column 4: amount of text messages (ATM) -- represents how many text messages a pair of nodes exchanged.
	
	\item column 5: friendship strength -- represents the information about the friendship between a pair of nodes participating in the experiment.
	
\end{itemize}

We model the possible values of these attributes as follows:

\begin{itemize}
	\item FM -- weak, average, high.
	\item CD -- weak, average, high.
	\item AC -- weak, high.
	\item ATM -- weak, high.
\end{itemize}

We model each attribute using the values from the MIT's experiment trace where, for example, FM is represented by numerical values (0 to 50 encounter times). Thus, we create three levels of FM, and convert these numerical values into FM levels. For example, from 0 to 4 encounters the frequency can be tagged as weak, from 5 to 15 it can be considered average, and over 15 it is considered high.

Additionally, to model the friendship strength between a pair of nodes, we have used the answers to one questions available in the MIT's experiment trace. The question is: "Is this person part of your close circle of friends?". The answer can be "yes" or "no". If node A answers "yes" to this question about node B, we assume in our training data 
that the friendship between A and B is \textit{strong}. If node A answers "no" to the question, the friendship is defined as \textit{weak}. Thus, the training data for the 
Naive Bayes algorithm consists of tuples with values relative to the aforementioned attributes, and to the friendship strength between nodes. Figure 1 presents an example
of the system used to classify the friendship strength between a pair of nodes in this work. Table \ref{tab:Table1} presents an example of some tuples from the training data used in this work.

\begin{algorithm}[ht]
	\caption{Step 1: Classification of friendship strength}\label{alg2}
	\begin{algorithmic}[1]
		\label{alg:the_alg}
		\Procedure{NaiveBayesClassifier}{$nodeA,nodeB$}
		\State $friendStrength \gets 0$
		\State $probIsStrong \gets calcProb(history Data From Nodes)$
		\State $probIsWeak \gets calcProb(history Data From Nodes)$
		\If{$probIsStrong > $probIsWeak}
		\State $friendStrength \gets $strong
		\Else
		\State $friendStrength \gets $weak
		\EndIf
		\EndProcedure
	\end{algorithmic}
\end{algorithm}

Next we provide more details concerning the selected Naive Bayes classifier.

\subsection{Naive Bayes classifier}\label{subsubsec:naive_bayes}

The Naive Bayes classifier uses the Bayes' theorem to calculate the probabilities necessary to classify a new instance. From a machine learning perspective, we can define 
the Bayes' theorem as follows: given an unknown instance $A = (a_1, a_2, a_3.. a_n)$, where $a_i (i=1, \ldots n)$ are the values of the attributes of an instance, we wish 
to predict to which class does $A$ belong to.

According to the Bayes' theorem, the probability of choosing a class given a sample is given by:

\begin{equation} \label{eq:1}
P(Class|A) = \frac{P(A|Class) \times P(Class)}{P(A)}
\end{equation}

We can rewrite equation \ref{eq:1} in terms of the attributes of instance A:

\begin{equation} \label{eq:2}
P(Class|A) = \frac{P(Class|a_1,...,a_n) \times P(Class)}{P(a_1,...,a_n)}
\end{equation}

The class definition of instance A involves the computation of the probability for all possible classes given a particular attribute. The defined class is the one with the 
highest probability. From a statistical perspective, this is equivalent to maximizing P(Class | $a_1$, Class | $a_2$, .., Class | $a_n$). The denominator in equation 
\ref{eq:2} is constant, which leads to the following simplification:

\begin{equation}  \label{eq:3}
argmax P (Class) \times \prod_{i}^{}  P(a_1,...a_n|Class)
\end{equation}

The next subsection presents an example of Naive Bayes classification.

\begin{table}
	\caption {A sample of the training data used in this work. This training data is generic for all the nodes.}
	\label{tab:Table1}      
	\begin{tabular}{lllll}
		\hline\noalign{\smallskip}
		\textbf{FM} & \textbf{CD}& \textbf{AC}& \textbf{ATM}& \textbf{Strength}\\
		\noalign{\smallskip}\hline\noalign{\smallskip}
		weak & no & no & high & weak\\
		weak & no & no & weak & weak\\
		average & no & no & high & weak\\
		high & yes & yes & high & weak\\
		high & yes & yes &  weak & strong\\
		high & yes & yes &  weak & strong\\
		average & yes & no & high & weak\\
		weak & no & yes & high & weak\\
		high & yes & yes &  weak & strong\\
		average & no & no &  weak & weak\\
		average & no & no & high & strong\\
		high & yes & yes & high & weak\\
		weak & yes & yes & high & strong\\
		average & yes & no &  weak & weak\\
		weak & yes & yes & high & weak\\
		\noalign{\smallskip}\hline
	\end{tabular}
\end{table}

\subsubsection{Example of Naive Bayes classification}\label{subsubsec:example}

We now present an example to clarify our approach. Consider that, at a given time, nodes A and B are in the coverage area, i.e., they can see each other; consider also that node B carries a message to node C. From the contact history of A and C we get the following instance \textit{{FM = high, CD = high, AC = no, ATM = yes}}. Given all this, FSF has to decide whether to forward the message to node A. To this end, FSF asks the Naive Bayes algorithm for the friendship strength between A and C. To answer this question, Naive Bayes follows the steps below:

\vspace{5mm}
\noindent\textbf{Step 1} -- Determine the probability of occurrence of each class:

\begin{equation}  \label{eq:4}
P(Class) = \frac{P}{Q}
\end{equation}

\noindent where $P$ is the number of cases of class $X$, and $Q$ is the total number of cases.

P(Strength = weak) = 10/15 = 0.66

P(Strength = strong) = 05/15 = 0.34

\vspace{5mm}
\noindent\textbf{Step 2} -- Determine the probability of the attributes for the instance in question about every possible class:

\begin{equation}  \label{eq:5}
P(Attribute(Z)|Class) = \frac{R}{S}
\end{equation}

\noindent where $R$ is the total number of cases of class $Y$ concerning characteristic $A_i$, $S$ is the number of cases of class $Y$, and $Z$ is the value for characteristic $A_i$.

P(FM = high AND strength = strong) = 2/5 = 0.4

P(FM = high AND strength = weak) = 3/10 = 0.3

P(AC = no AND strength = strong) = 2/5 = 0.4

P(AC = no AND strength = weak) = 4/10 = 0.4

P(ATM = yes AND strength = strong) = 4/5 = 0.8

P(ATM = yes AND strength = weak) = 4/10 = 0.4

P(CD = high AND strength = strong) = 4/5 = 0.8

P(CD = high AND strength = weak) = 5/10 = 0.5

\vspace{5mm}
\noindent\textbf{Step 3} -- Determine the probability of each class based on the probability of the instance:

\begin{equation}  \label{eq:6}
P(Class|a_1,a_2,...,a_n) = \prod P(a_i|Class) \times P(class)
\end{equation}

\begin{equation}  
P(A|class = strong) = (0.4 \times 0.4 \times 0.8 \times 0.8 \times 0.34) = 0.034
\end{equation}

\begin{equation}  
P(A|class = weak) = (0.3 \times 0.4 \times 0.4 \times 0.5 \times 0.66) = 0.015
\end{equation}

In this example, and according to the calculated probabilities, the friendship strength 
between nodes A and C would be classified as \textit{strong}.

\begin{figure*}
	\centering
	\begin{subfigure}[htb]{0.24\textwidth}
		\includegraphics[width=\textwidth]{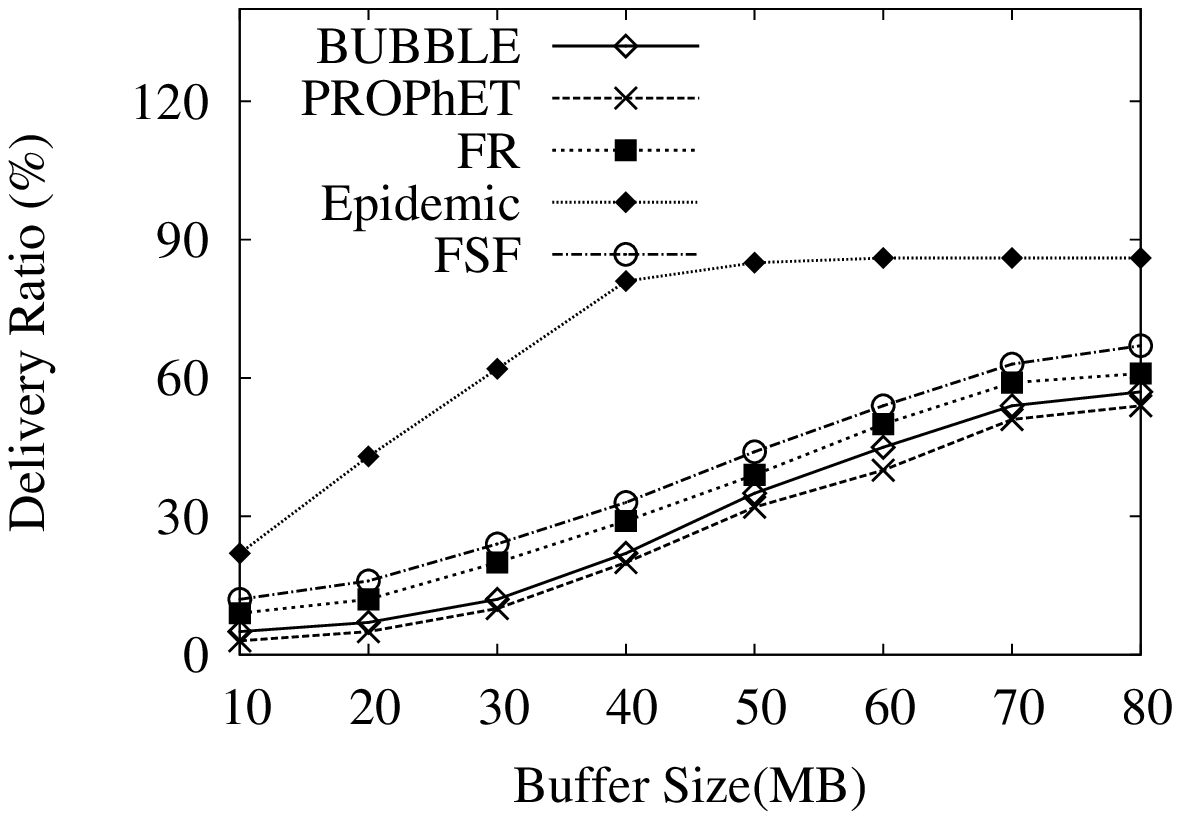}
		\caption{Delivery ratio}
	\end{subfigure}
	\begin{subfigure}[htb]{0.24\textwidth}
		\includegraphics[width=\textwidth]{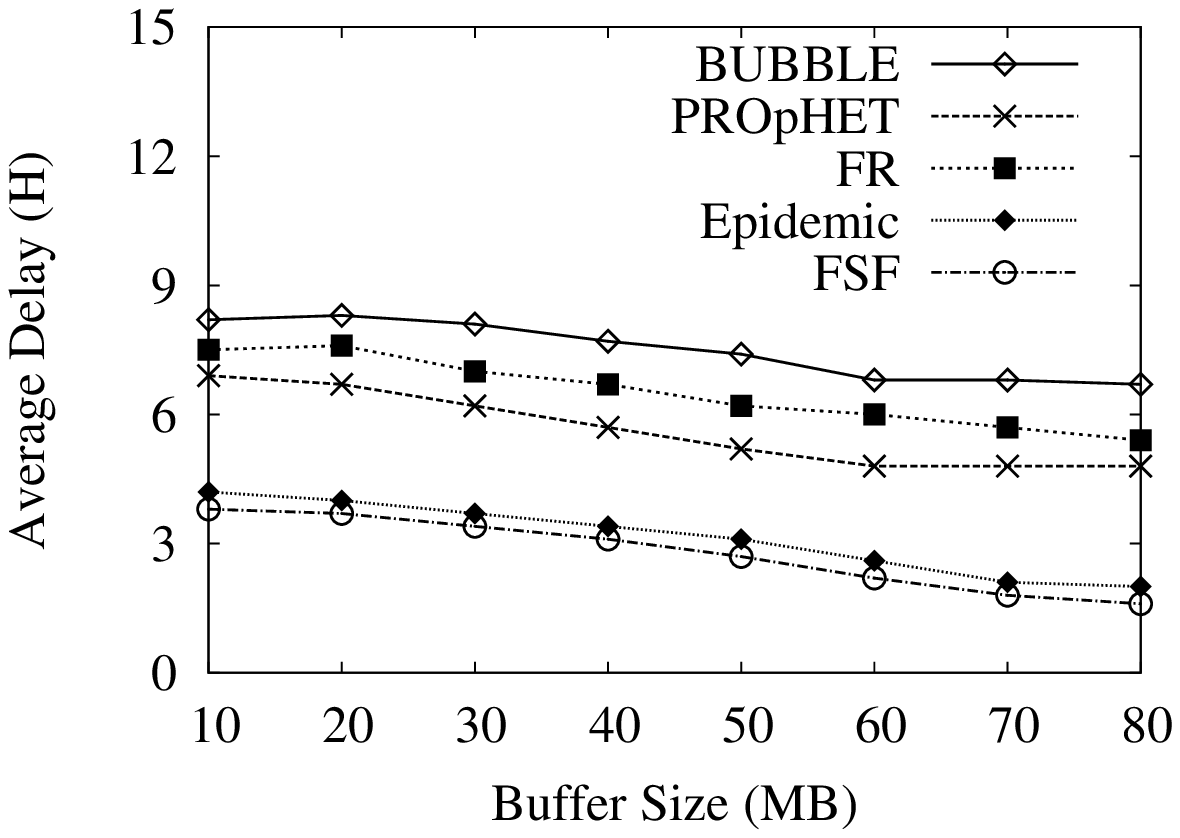}
		\caption{Average Delivery Delay}
	\end{subfigure}
	\begin{subfigure}[htb]{0.24\textwidth}
		\includegraphics[width=\textwidth]{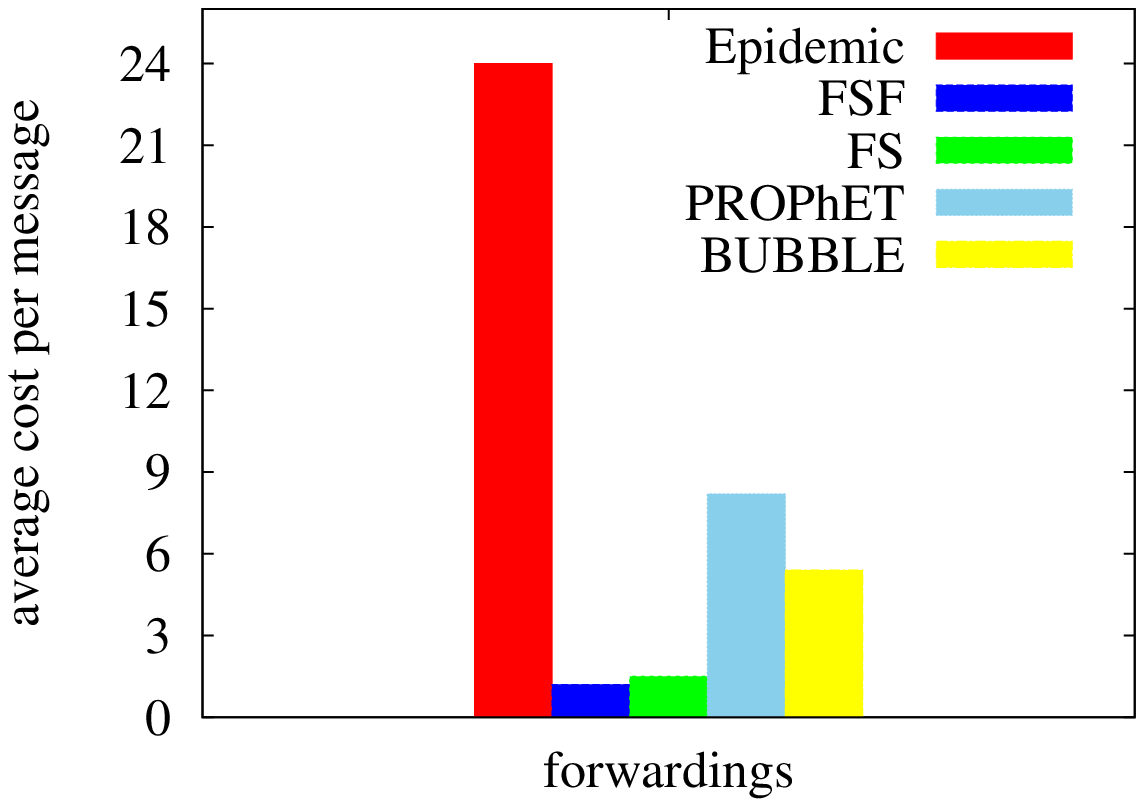}
		\caption{Average Cost}
	\end{subfigure}
	\begin{subfigure}[htb]{0.24\textwidth}
		\includegraphics[width=\textwidth]{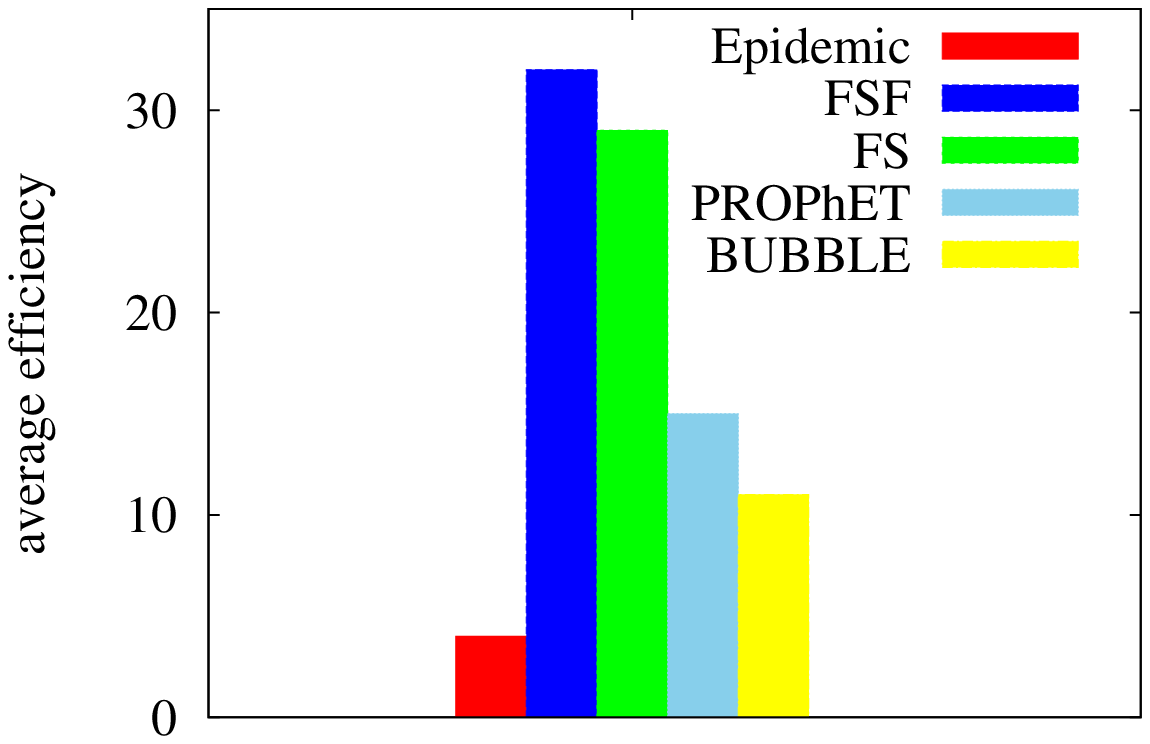}
		\caption{Average Efficiency}
	\end{subfigure}
	\caption{Simulation results for the Cambridge scenario.}
	\label{fig:Figure1}
\end{figure*}

\subsection{Step (ii): Selfish node detection}\label{subsec:classification_selfishness}

After evaluating the friendship strength, FSF has to classify the selfishness from the message relay point view. In this work, and from a selfishness perspective, a network node
can be classified in three ways: (i) individually selfish, (ii) socially selfish and (iii) not selfish. Below, we provide more details about each of these types of nodes.

\begin{itemize}
	\item \textbf{(i) individually selfish} -- according to the selfishness classification made in \cite{miao2013investigation}, individual selfishness can be defined as ``the unwillingness of a single node to relay the messages of all other nodes in order to conserve its limited resources''. These nodes basically do not accept messages.
	
	\item \textbf{(ii) socially selfishness} -- can be defined as ``a type of selfishness in which a node belongs to a community, and it is willing to relay messages for the nodes 
	within the same community, but refuses to relay messages for the nodes outside its community \cite{miao2013investigation}''. So, these nodes will only relay messages for nodes belonging to his/her community.
	
	\item  \textbf{(iii) not selfish} -- those nodes accepting to receive messages to all other nodes. However, we have considered that non-selfish nodes can behave in a 
	selfish manner in situations where he/she cannot handle more messages due to resource depletion. 
	In such cases, the message relay will not accept to receive messages to save its resources. More details will be given in subsection \ref{not:selfish}
	
\end{itemize}

\begin{algorithm}[ht]
	\caption{Step 2: Node selfishness assessment}\label{alg3}
	\begin{algorithmic}[1]
		\label{alg:the_alg}
		\Procedure{SelfishnessAssessment}{$node$}
		\State $reputationLevel \gets hisReputation(node)$
		\If{$reputationLevel > $0.7}
		\State $node.reputation \gets $Selfish
		\Else \If{$memoryLevel > \alpha  OR  ennergyLevel < \beta $}
		\State $node.reputation \gets $Selfish
		\EndIf
		\EndIf
		\EndProcedure
	\end{algorithmic}
\end{algorithm}

Algorithm \ref{alg3} presents the classification of the node selfishness procedure. To classify node selfishness, FSF uses a reputation system based on \cite{soares2014statistical}. Below, more details will be provided. 

\subsubsection{Reputation system}

In this work, to classify the node selfishness, we have used a reputation system based on \cite{soares2014statistical}.  
It was implemented in the hisReputation() method presented in Algorithm \ref{alg4}. In our reputation system, the network has a set of $N$ mobile nodes and $S$ selfish nodes. 
Initially, the nodes have the same reputation, and no information about the network. Figure \ref{fig:1} shows the selfishness classification
module running at each network node. Each node operates in a promiscuous mode, and it may listen to every bundle transmitted by its neighbors. All the nodes have a common 
technique to detect selfish behavior by overhearing the bundles transmitted and received by its neighbors, in order to detect anomalies. 

To address it, each node has two main components acting with the network interface. The first one is the detection scheme, responsible of monitoring the data traffic 
when searching for selfish nodes, and the second one is the reputation system, which measures the cooperation level of its neighbors. 

Each node carries a monitoring module to assess the behavior of neighbor nodes in the network. Upon each contact, some assessment is performed no matter whether a node is selfish or not. Thus, node A can get information from a neighbor B when one of the following situations occurs:

\begin{description}
	\itemsep0em
	\item[\textbf{Selfish Contact}] \hfill \\
	Node A, using its monitoring mechanism during a contact, detects that neighbor node B is selfish. 
	
	\item[\textbf{Not Selfish Contact}] \hfill \\
	Both nodes A and B are not selfish. So, they may share information about individual probes to feed the distributed reliable system. 
	
\end{description}

We model the contact between the nodes of the network as an infinitely repeated game. Each contact is a game $J = (N, A)$ with $N$ players (nodes), being A = \{cooperate,
do not cooperate\}, the set of possible actions of a player during an interaction. During a contact, node $u$ wants node $v$ to forward its message over the 
network, but $v$ may or may not have interest in cooperating based on the trade-off between benefit and cost ratio related to the transfer. Thus, we define a round involving
a pair of nodes in the contact $(u, v) \in E$ as the pair $(A_u, A_v)$ employed by the pair $(u, v)$ during the contact.

All non-selfish nodes are able to detect if a neighbor node is selfish. Each node can maintain a table of type $ (ID, R) $ which
serves to maintain the individual history on its neighbors and assign a reputation value $R$ for each node $ID$ in accordance with the observations. We
call $R$ the degree of collaboration associated with the contact between $(u, v)$.

Even for collaborative nodes, reputation can be discriminating because of the social process associated with opportunistic networks. Collaborative nodes with a stronger
social bond will have a higher reputation than collaborative nodes with weaker social ties. To address this variation of the problem, we use a variant of the sigmoid
function \cite{rodriguez2003analytical}, a mathematical strategy commonly applied in the literature on learning systems and neural networks, among others. Sigmoidal 
functions are able to assess the probability of an event to occur based on system observations. When the network is started, the cooperation probability
is the same for all nodes, but as the observations related to cooperation are collected from the network, these values are adjusted until cooperation probability reaches a 
value with little variability. This situation is called mature stage, when we got the minimum number of observations to achieve a stable learning situation. When a 
behavior on the network is changed, like a node turning selfish, the sigmoidal functions lead to a learning stage until it again reaches a mature stage.

The observations regarding the behavior of neighbor nodes become sufficient at some point in time to accurately understand a particular state of the system. The understanding of 
the systems becomes reliable even under small variations in the observations. These variations can raise from communication errors, errors in the detection system, and even 
when network nodes change their behavior. These variations can raise from communication errors and errors in the detection system. We apply this function to the reputation 
model from new observations collected from the network until the ranking can be considered reliable, with little risk of false negatives or false positives.

Since in opportunistic networks the total number of nodes is unknown, the scalability has to be considered, as well as the variation of the cooperation level due to reaction
methods as methods of punishment/incentive. Thus, we discuss methods of further reaction with no major disruption in the reputation data. Another remarkable feature of 
these functions is the display of learning curves that can be used to predict the future behavior. This is a valuable function, as we can predict selfish nodes with more 
accuracy and less information. This way, we analyze the probability of a node $u$ to be more cooperative than $v$ as a decision process defined by Eq.~(\ref{eq:eq1}):

\begin{equation} \label{eq:eq1}
P_{cooperation}(u,v) = \frac{1}{1 + 10^{\frac{R_v-R_u}{F_d}}}
\end{equation}

where $P_{cooperation}(u, v)$ describes the probability of $u$ be more cooperative than $v$, $R_k$ is the reputation of node $k \in N$, and $F_d$ is a significance factor to
stress the difference between the reputations of pair $(u, v)$. Assuming $F_d = 5$, and the reputation of the pair $(R_u, R_v) = (10,4)$, as well as $| R_v - R_u | > F_d$, 
the variation is significant, evidencing that node $u$ is much more cooperative than $v$, $P_{after_cooperation}(u, v) \simeq 0.94$. However, if $(R_u, R_v) = (8,7)$, 
then $P_ {cooperation}(u, v) \simeq 0.61$, which means that the difference between reputations may not be significant for an accurate assessment of the cooperation level
between them, whereas in the first case the difference may mean that $v$ can be selfish. 

Thus, we define the update process as a two-step process for each pair of contacts $(u, v)$: update the reputation for node $v$ and the reputation of the neighbor nodes of
$u$, except $v$. In this case, we choose to create the second step as a way of encouraging non-selfish nodes by increasing their personal reputation through positive for
selfish nodes. This update, however, only occurs when there is a positive selfishness on node $v$, not penalizing other nodes (selfish and not selfish) when in contact
with $v$, a non selfish node. The process of reputation updating  is the following:

\begin{description}
	\item[\textbf{Average reputation}] \hfill\\
	Upon each contact, the node calculates its average reputation using
	Eq.~\ref{eq:eq3}, the arithmetic mean of the reputation of the neighbors at the time of 
	contact with node $v$.
	\begin{equation} \label{eq:eq2}
	AR(u) = \frac{\sum_{\forall k \in neighborhood_u - \{v\}} R_k}{|neighborhood_u| - 1}
	\end{equation}
	\item[\textbf{Node update}] \hfill\\
	The individual update of a node, when there is a positive for a selfish node,
	is computed through Eq.~\ref{eq:eq2}, where $D (v) \in \ {0,1} $ assumes 0 when $ v $ is selfish,	and 1 otherwise. $\delta$ is the weight assigned to each new observation.
	
	\begingroup
	\setlength\abovedisplayskip{0pt}	
	\begin{equation} \label{eq:eq3}
	R_v{}^{'} = R_v + RF(D(v) - P_{cooperation}(u,v))
	\end{equation}
	
	\item[\textbf{Neighbor update}] \hfill\\
	The update of the other neighbor nodes is given by Eq.~\ref{eq:eq2},
	with $D(k) = 1$, $P_{cooperation}(k, v) \forall k \in neighborhood_u$.
	\endgroup
\end{description}

Overall, by using the reputation model, each node can infer the behavior of a neighbor node based on its reputation value. Therefore, we introduced a verification algorithm which allows the classification of the behavior of a neighbor node by collecting a number of observations. For this purpose, we use a clustering algorithm to classify the 
observed reputations in two classes: selfish and non-selfish neighbors. Since reputation is a $float$ type of value, we also can use this algorithm to classify how well the two clusters are separated. This information is also important to avoid wrong classifications due to errors on the detection process. For this purpose, we use a clustering algorithm
to choose the moment beyond which available samples are enough to distinguish between the reputations of selfish and non-selfish nodes. Algorithm \ref{alg4} presents the
reputation updating method executed by each network node.

\begin{algorithm}[H]
	\begin{algorithmic}[1]
		\Procedure{HisReputation}{$v$}
		\State $D \longleftarrow 1$ \Comment{When a behavior detection of v occurs on node u}
		\State $R_{y} \longleftarrow {\frac{ \sum_{\forall k \in neighborhood_{u} }R_{k} }{ \left |neighborhood_{u}-1 \right | }}$
		\State $probCoop(y,v) \longleftarrow 1$
		
		\If{ \emph{v was detected as selfish}}
		\State $D \longleftarrow 0$
		\EndIf
		
		\State $probCoop(y,v) \longleftarrow {\frac{1}{1 + 10^{\frac{R_{v} - R_{y}}{F_{d}}}}}$ \Comment{Calculate the probability of cooperation of v in relation to the other neighbors of u}
		
		\State $R_{v} \longleftarrow {R_{v} + \delta (D - probCoop(y,v))}$ \Comment{Update the reputation of v on the reputation table of u}
		
		\caption{Update Reputation Table.}\label{alg4}
		\EndProcedure
	\end{algorithmic}
\end{algorithm}

\subsubsection{Non-selfish nodes}\label{not:selfish}

The FSF algorithm also checks the third aforementioned situation where a non-selfish node behaves as a selfish node because his/her resources are at critical levels. In this 
work, we propose a model that takes into account two fundamental resources: battery level and memory availability. When an opportunity for a contact arises, and the friendship strength among the nodes is considered strong, FSF checks the selfishness of the message relay to decide whether to forward the message. Hence, any message will be forwarded if, and only if, the available resources at the message relay device have not reached critical levels.

From the example in section \ref{subsec:forwarding_strategy}, the friendship strength among nodes A and C was classified as strong. However, assume that the device carried 
by node A has only 50\% of battery energy level, and/or up to 70\% of used memory space. In such a case, it is reasonable to assume that node A, the candidate node to relay the 
message, does not accept the message to node C because it has to save its own resources.

\begin{figure*}
	\centering
	\begin{subfigure}[htb]{0.24\textwidth}
		\includegraphics[width=\textwidth]{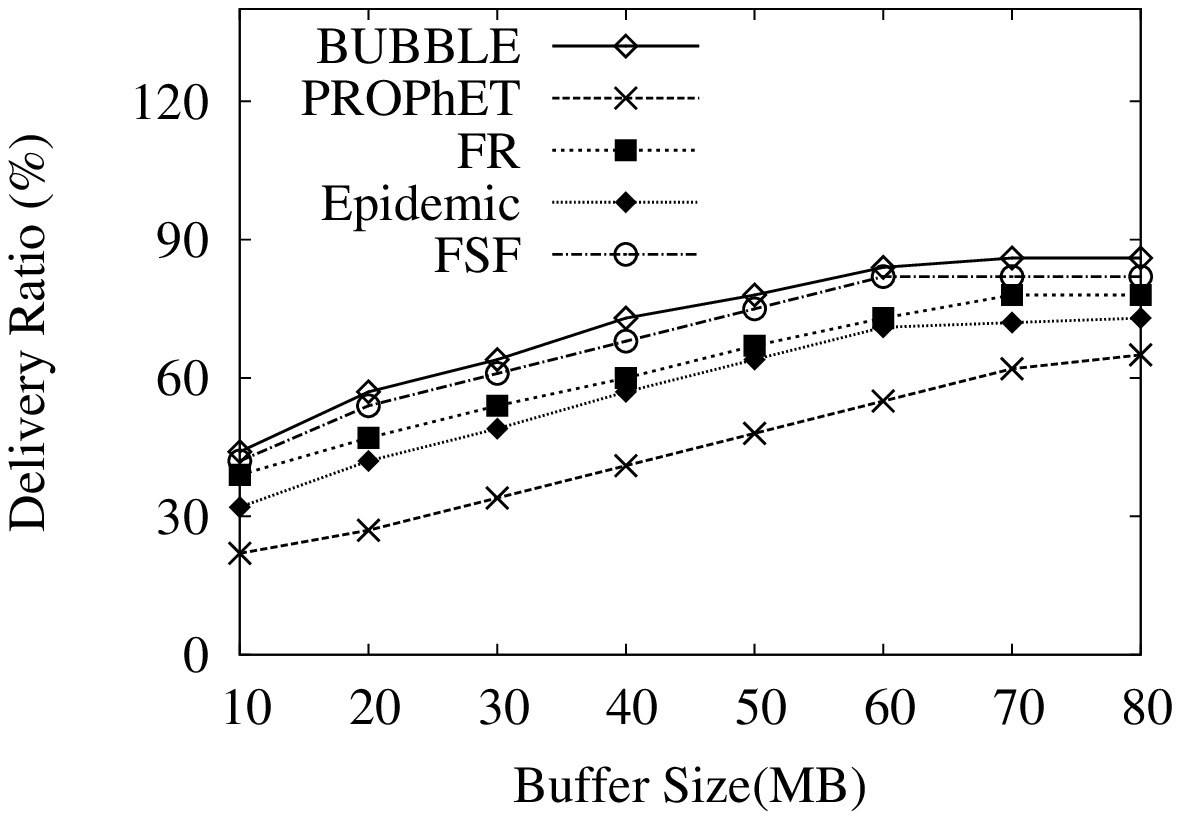}
		\caption{Delivery ratio}
	\end{subfigure}
	\begin{subfigure}[htb]{0.24\textwidth}
		\includegraphics[width=\textwidth]{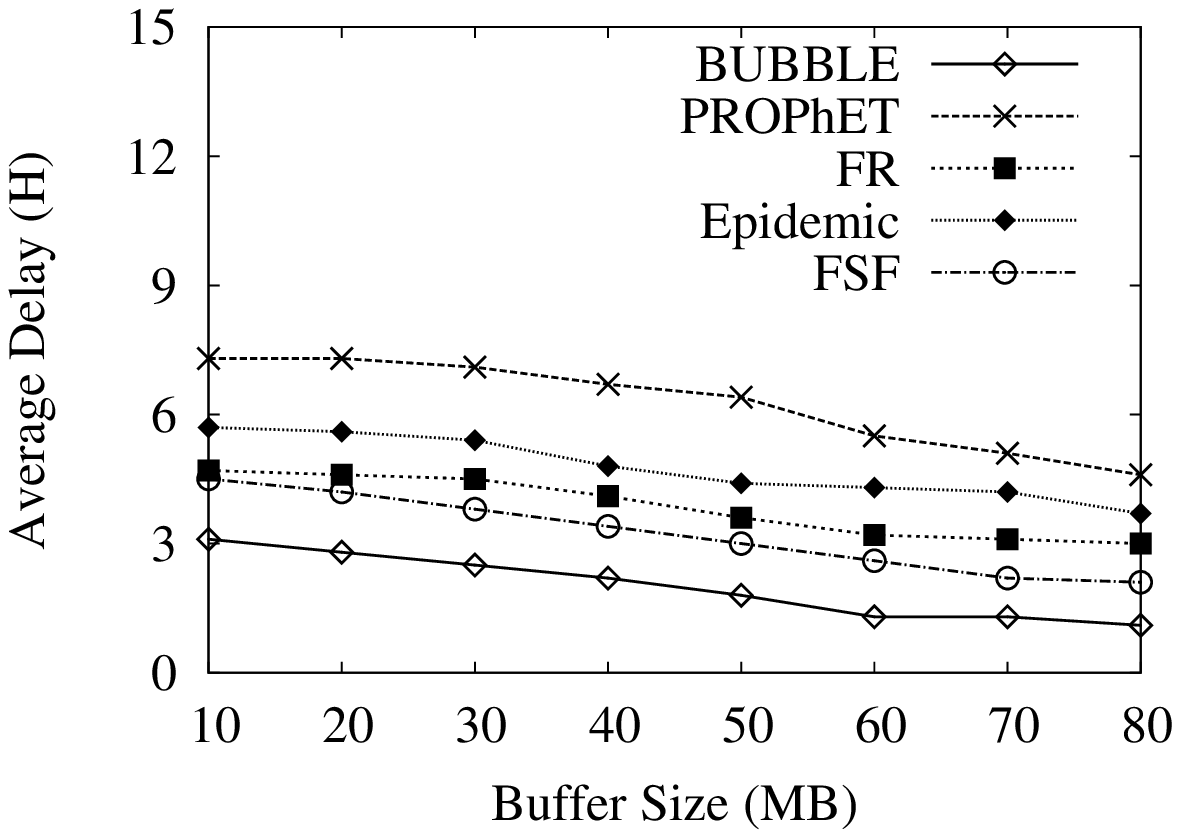}
		\caption{Average Delivery Delay}
	\end{subfigure}
	\begin{subfigure}[htb]{0.24\textwidth}
		\includegraphics[width=\textwidth]{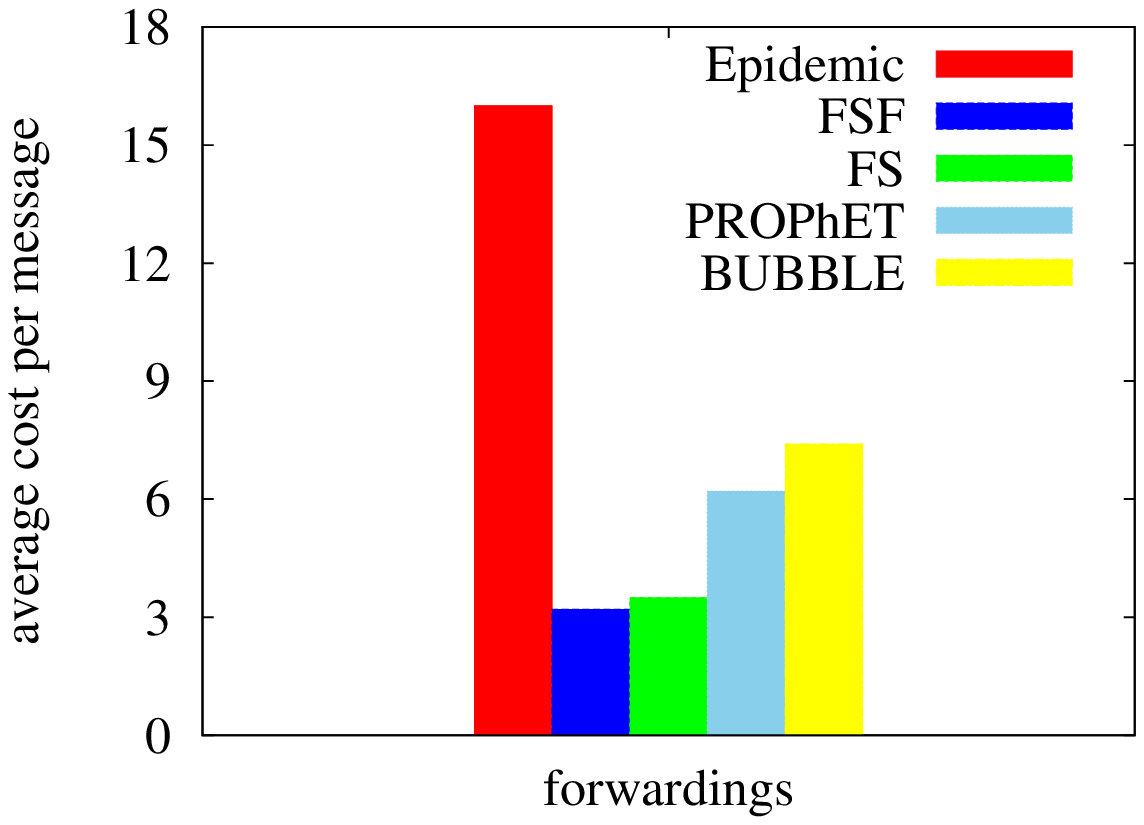}
		\caption{Average Cost}
	\end{subfigure}
	\begin{subfigure}[htb]{0.24\textwidth}
		\includegraphics[width=\textwidth]{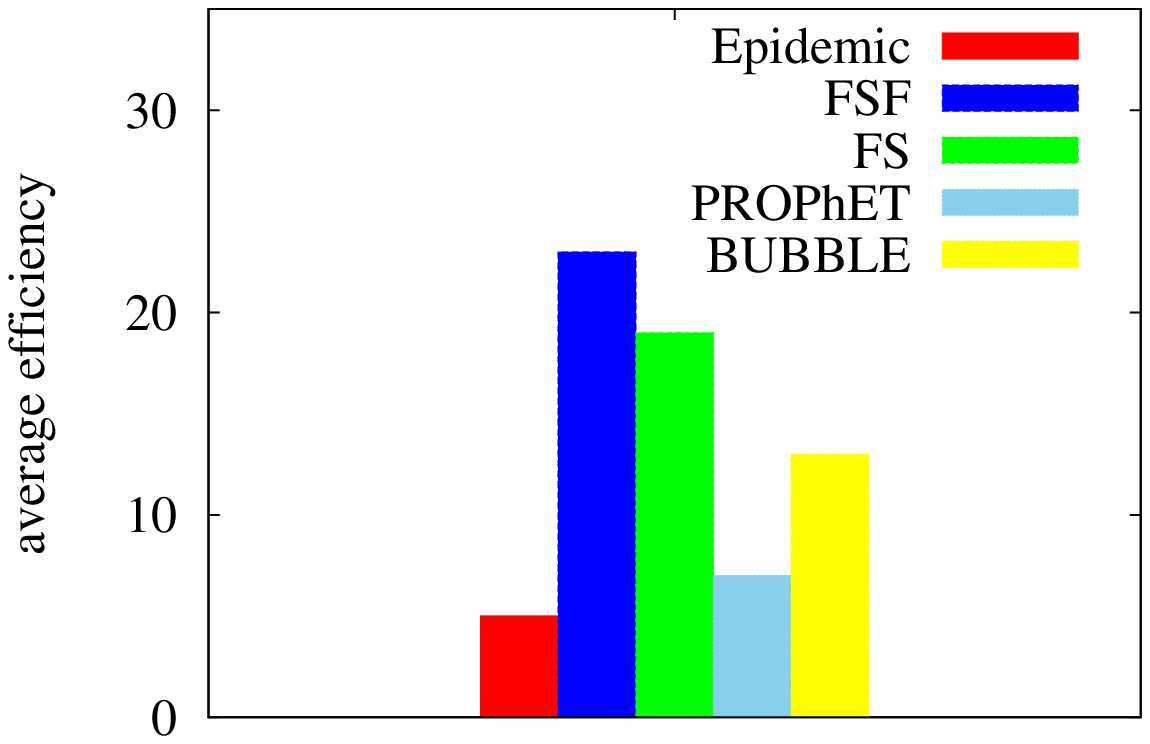}
		\caption{Average Efficiency}
	\end{subfigure}
	\caption{Simulation results for the Infocom05 scenario.}
	\label{fig:Figure2}
\end{figure*}

\begin{figure*}
	\centering
	\begin{subfigure}[htb]{0.24\textwidth}
		\includegraphics[width=\textwidth]{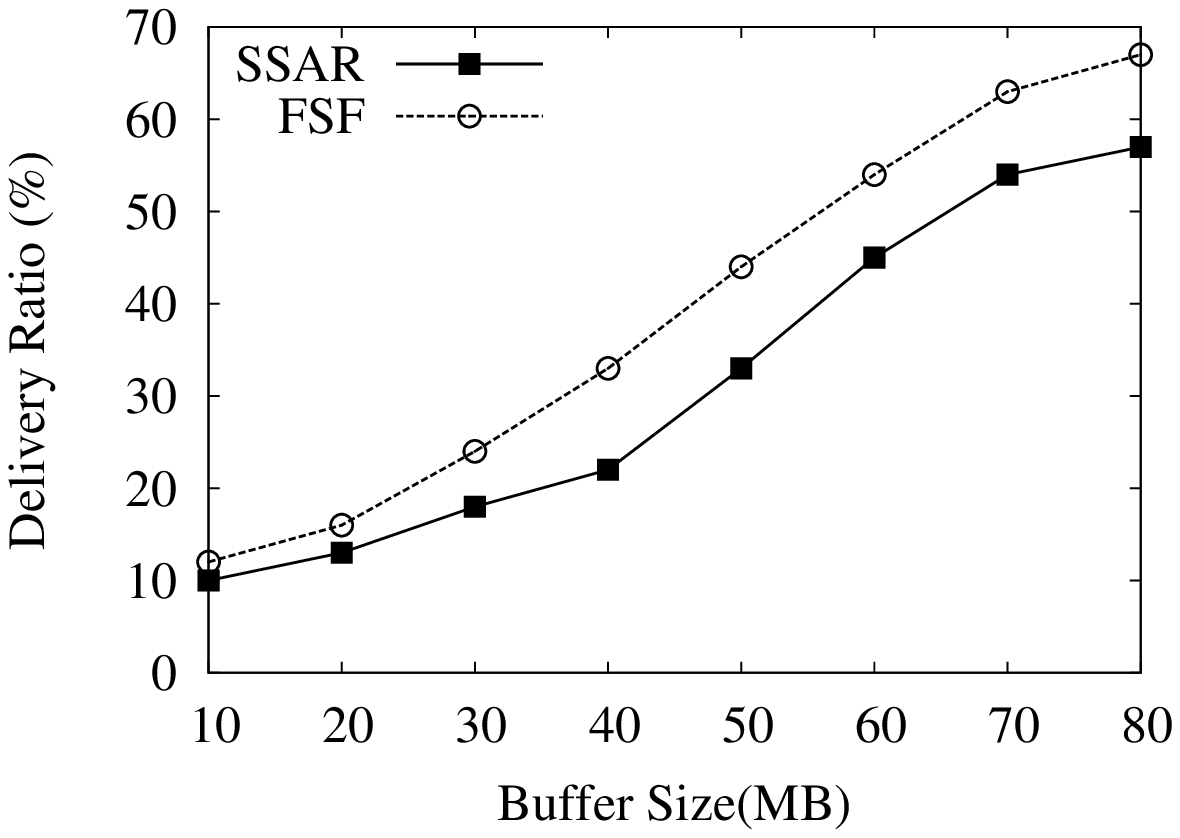}
		\caption{Delivery ratio}
	\end{subfigure}
	\begin{subfigure}[htb]{0.24\textwidth}
		\includegraphics[width=\textwidth]{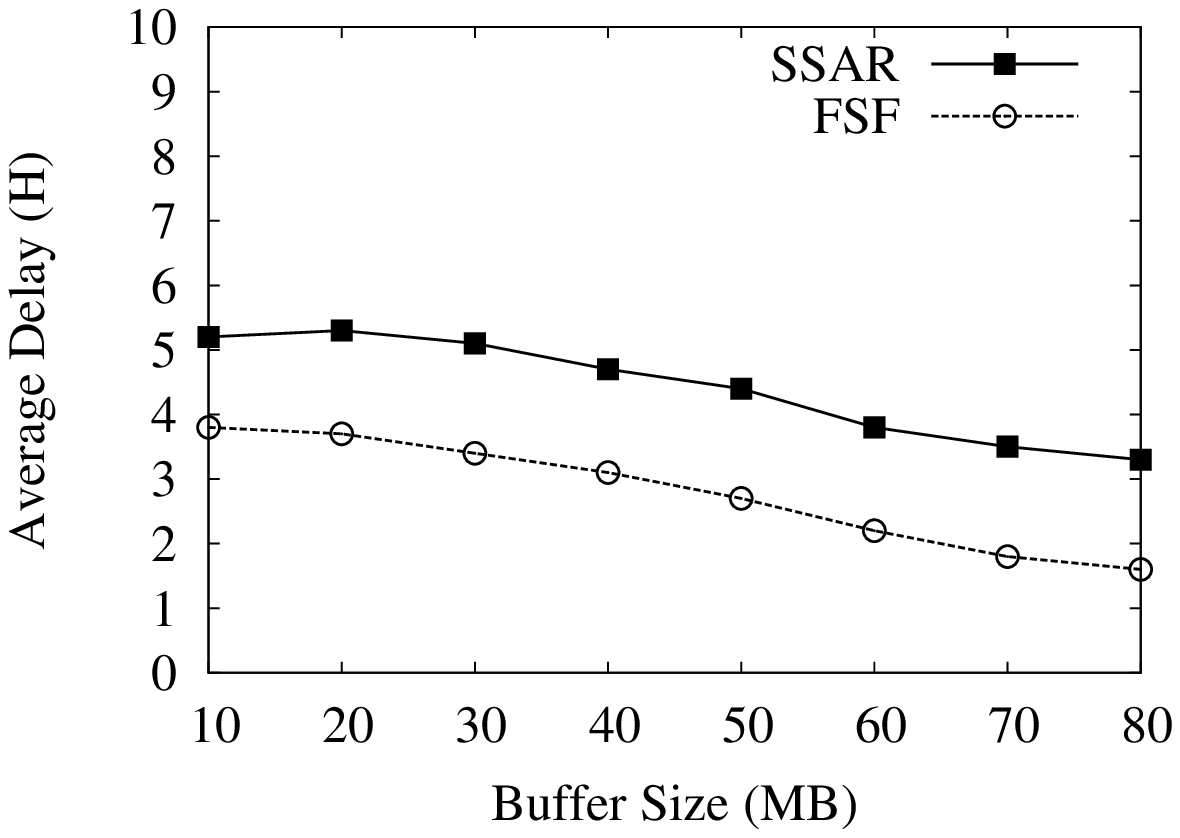}
		\caption{Average Delivery Delay}
	\end{subfigure}
	\begin{subfigure}[htb]{0.24\textwidth}
		\includegraphics[width=\textwidth]{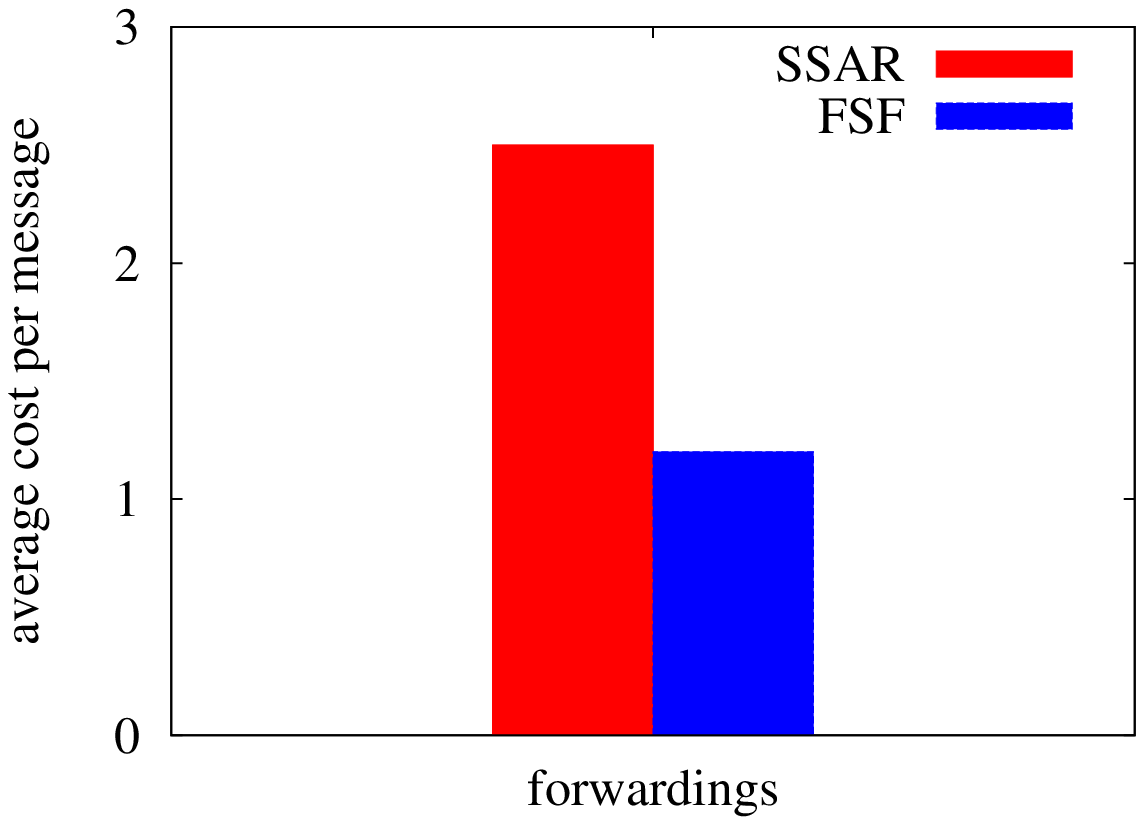}
		\caption{Average Cost}
	\end{subfigure}
	\begin{subfigure}[htb]{0.24\textwidth}
		\includegraphics[width=\textwidth]{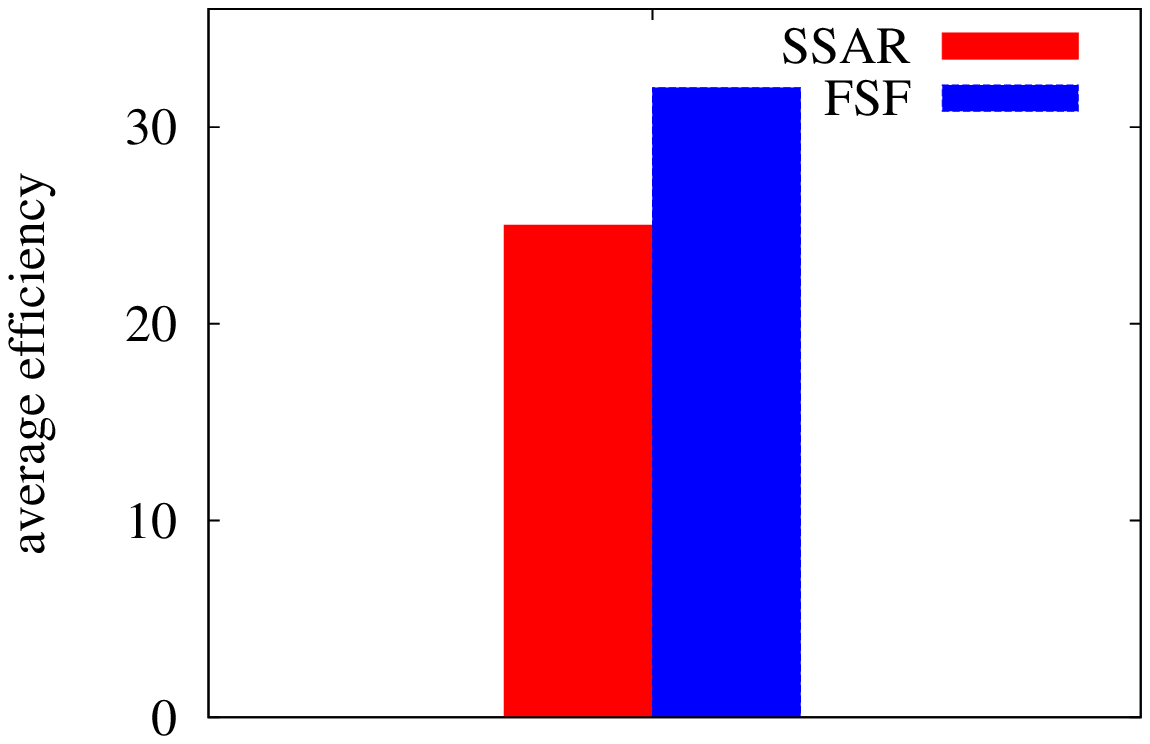}
		\caption{Average Efficiency}
	\end{subfigure}
	\caption{Comparison of FSF x SSAR - simulation results for the Cambridge scenario.}
	\label{fig:Figure3}
\end{figure*}

To simulate this situation we consider that, when a contact opportunity arises, if the message relaying device has an energy level less than $\alpha$, or if the used memory 
space is more than $\beta$, the node will decline to relay the message, although the friendship among the corresponding nodes may be strong. For the sake of experimentation, in our simulations, we consider $\alpha=30\%$ and $\beta=70\%$.

\section{Evaluation methodology}\label{sec:evaluation}

In this section, we present the selected evaluation methodology to evaluate the performance of our FSF proposal.

Table~\ref{tab:Table3} and Table~\ref{tab:Table2} list the simulation parameters used in our simulation experiments. Messages were generated every two seconds, when a new pair of source/destination nodes were chosen randomly.

\begin{table}
	\caption{Some parameters deployed in the simulation.}
	\label{tab:Table3}
	\begin{tabular}{ll}
		\hline\noalign{\smallskip}
		\noalign{\smallskip}\hline\noalign{\smallskip}
		\textbf{Transmission speed} & 250 kbps\\
		\textbf{Message generation rate} & 2 msg/sec\\
		\textbf{Message size} & 0,5 - 1 MB\\
		\textbf{TTL} & 5 h\\
		\noalign{\smallskip}\hline
	\end{tabular}
\end{table}

\subsection{Experimental setup}

To evaluate our FSF algorithm, we performed trace-driven simulations using the ``ONE Simulator'' \cite{keranen:2009}. To simulate node mobility, we used two well known 
real mobility traces: Cambridge and Infocom05 traces, respectively. In the Cambridge trace, the devices were distributed to students from the University of Cambridge, mainly to students of the Computer Laboratory, specifically PhD and Master students and undergraduates. 

\begin{table}
	\caption{Some parameters about traces utilized in the simulation.}
	\label{tab:Table2}
	\begin{tabular}{lll}
		\hline\noalign{\smallskip}
		\noalign{\smallskip}\hline\noalign{\smallskip}
		\textbf{Trace} & \textit{Cambridge} & \textit{Infocom05} \\
		\textbf{Device} &\textit{iMotes} & \textit{iMotes}\\
		\textbf{Network Interface}  &\textit{Bluetooth} &\textit{Bluetooth} \\
		\textbf{Nodes} & 54 & 41 \\
		\textbf{Trace Duration} & 11 days & 4 days\\
		\textbf{Granularity} & 100s & 120s\\
		\textbf{Number of contacts} & 10873 & 22459\\
		\noalign{\smallskip}\hline
	\end{tabular}
\end{table}

In the Infocom05 trace, the iMotes were distributed to students taking part while attending Infocom workshops. Despite these students arrived from different countries, they all attended the same event, and/or most of them stayed in the same hotel for 4 days. Table \ref{tab:Table2} provides more details about these two mobility scenarios.

\subsection{Routing algorithms}

We implemented a simulation model which includes FSF and, for the sake of performance comparison, we also implemented four well-known algorithms found in the literature: 
BUBBLE-RAP \cite{hui:2011}, Friendship routing (FS) \cite{bulut2010friendship},
PROPhET \cite{lindgren:2003} and Epidemic \cite{vahdat:2000}. 

The BUBBLE-RAP algorithm forwards messages by taking into account node popularity and its community information.
We selected this algorithm since it is well known in the DTN literature, and also because it adopts a strategy that accounts for social characteristics. The FS algorithm 
forwards messages by taking into consideration the value of a metric called Social Pressure Metric (SPM), based on characteristics like frequency of contacts, longevity, 
and regularity.  FS is considered the first routing mechanism based on the friendship level among DTN nodes. The PROPhET routing algorithm uses the history of contacts and 
transitivity in routing decisions. When two nodes meet they exchange summary vectors, which in this case also contain the delivery predictability information stored at the 
nodes. This information is used to perform the routing decisions. The operation of the Epidemic routing protocol is similar to that of PROPhET, but in a contact opportunity the nodes exchange summary vectors containing information about the messages stored in their buffer. Thus, they exchange all the messages stored in their respective buffers. Both 
algorithms, PROPhET and Epidemic, are selected because they are the most widely used routing algorithms in DTN routing comparisons.

\subsection{Node resource management}

Due to the constraints in terms of DTN node capacity, we had to select a buffer 
management policy to deal with overflow conditions. To avoid favoring the 
routing decisions of the algorithms under comparison, we selected the Drop 
Oldest buffer management strategy, which discards the oldest message in the 
case of buffer overflow.

To simulate the power consumption of nodes, we adopted the energy model proposed in \cite{silva2012energy}. The power consumption of a node is classified into five states:

\begin{itemize}
	
	\item Off -- no power consumption as the network node interface is turned off.
	\item Inactive -- reduced power consumption as the network node interface is idle.
	\item Scan -- the node consumes power while the network node interface detects neighbors.
	\item Transmission -- the node consumes power while sending a message.
	\item Reception -- the node consumes power because it is receiving a message.
	
\end{itemize}

In our simulations, we assume that all the nodes initially have the maximum level of energy. We consider that the energy level of the nodes is measured in units, being 500 units the highest value. We assume that the user recharges his/her device every 24 hours. Energy consumption depends on the device's state and the number of operations using the network interface. For example: if the node is at transmission, reception or scan states, we assume a reduction of 25 energy units for every sent/received message and/or for every scan performed. If the node is at off or inactive states, we assume there is no energy cost.

\subsection{Metrics}

The DTN performance can be measured in terms of average delivery ratio, and 
average delivery delay \cite{fathima:2008}. Moreover, we applied two other 
metrics widely used in other works found in the literature, i.e.
\cite{bulut2010friendship}, \cite{li:2010} and \cite{qin2015nfcu}:

\begin{itemize}
	\item Average delivery ratio -- the ratio of messages received by the destination nodes to those generated by the source nodes.
	
	\item Average delivery delay -- the average time interval among the sending and receiving events for messages traveling along the network.
	
	\item Average cost -- the average number of forwarding events per message delivered to the destination.
	
	\item Efficiency -- the ratio between delivery ratio and average cost.
\end{itemize}

\section{Results}\label{sec:results}

In this section, we present and discuss the results, as well as some important issues of our proposal such as the impact of individual selfishness on routing, and the classification analysis by a machine learning algorithm.

\subsection{Results for the Cambridge scenario }\label{subsec:delivery_rate}

Figure~\ref{fig:Figure1} presents the results for the delivery ratio obtained in the Cambridge scenario.
For the sake of comparison, we followed the strategy adopted in other works found in the literature, thereby varying the buffer size from 10 to 80 MB \cite{fathima:2008,krifa:2008,li:2009a}.

We find that the Epidemic algorithm outperforms the other ones by delivering more messages as the scenario characteristics potentially favor its strategy. As mentioned before, the Cambridge trace used two groups of undergraduate students (year 1 and year 2), as well as Masters and Ph.D. students from the Cambridge Computer Laboratory, running over 11 days. As these students belong to the same place, each message created will be sent to all other nodes belonging to the network, and it will increase the message delivery probability. Despite this, the Epidemic strategy can be very wasteful of network resources because it creates a lot of contention for the limited buffer space and network capacity, resulting in significant resource constraints \cite{lindgren:2003,spyropoulos:2005,tseng2002broadcast}.

The FSF algorithm achieves the second best result concerning to the delivery ratio, but we can see that the FS algorithm provides very similar performance levels. This occurs because both algorithms take friendship strength into account for their routing decisions. As referred above, we can suppose from the real world that, if the message relay and the message destination have a strong friendship strength, they have more probability of delivering messages to each other mainly because of the number of encounters. These two results also confirm the hypothesis raised in section 3.1 about the use of friendship strength among a pair of nodes as an appropriate criterion to use in DTN message routing. The difference between the performance of FSF and FS algorithms can be explained by the buffer management decisions. 

Buffer management can be defined as the strategy to choose which messages to drop when facing buffer overflow, and it is considered an important issue in DTNs according to \cite{qin2015nfcu}, as it can solve the negative effects of a low delivery ratio causing important messages to be deleted. For buffer management decisions, the FSF algorithm uses the Drop Less Known (DLK) approach, which drops the message addressed to a node with whom he/she has a weak social relationship, while the FS algorithm uses the First In First Out (FIFO) policy, which drops the first message stored in the buffer. Thus, the DLK policy maintains in the buffer messages that the FSF algorithm can deliver in future contacts, while FIFO can drop it from the buffer if the first message stored is addressed to a node with whom he/she has a strong friendship, and hence more delivery probability. Another important reason explaining why FSF outperforms FS is how it classifies the friendship strength among nodes. In this work, we proposed an approach for classifying the friendship strength using a machine learning algorithm. This algorithm has two important tasks: first, to learn what friendship is based on data collected from the real world. The second task is to classify new relationships among two nodes in the network. From the FSF results in the Cambridge scenario, it is reasonable to assume that Naive Bayes offers an excellent performance at classifying the friendship among nodes. So, the better the classification made by the Naive Bayes algorithm, the better the FSF performance.

Figure~\ref{fig:Figure1}.(b) presents the results for the average delivery delay in the Cambridge scenario. In DTN, a long delay can take hours or days. Thus, it is important to minimize the average delivery delay in DTN environments.

From the results shown in the Figure~\ref{fig:Figure1}.(b), we can see that Epidemic routing achieve the best result in terms of delivery delay, but we can see that, most of the time, FSF achieves an average delivery delay similar to Epidemic. Once again, we can justify this result by considering the scenario characteristics, that potentially favor the strategy of the Epidemic algorithm for the reasons referred above.
It is worth mentioning that, again, FSF outperforms the FS algorithm, and that the
BUBBLE and PROPhET algorithms achieved the lowest performance levels.

\begin{figure*}
	\centering
	\begin{subfigure}[htb]{0.24\textwidth}
		\includegraphics[width=\textwidth]{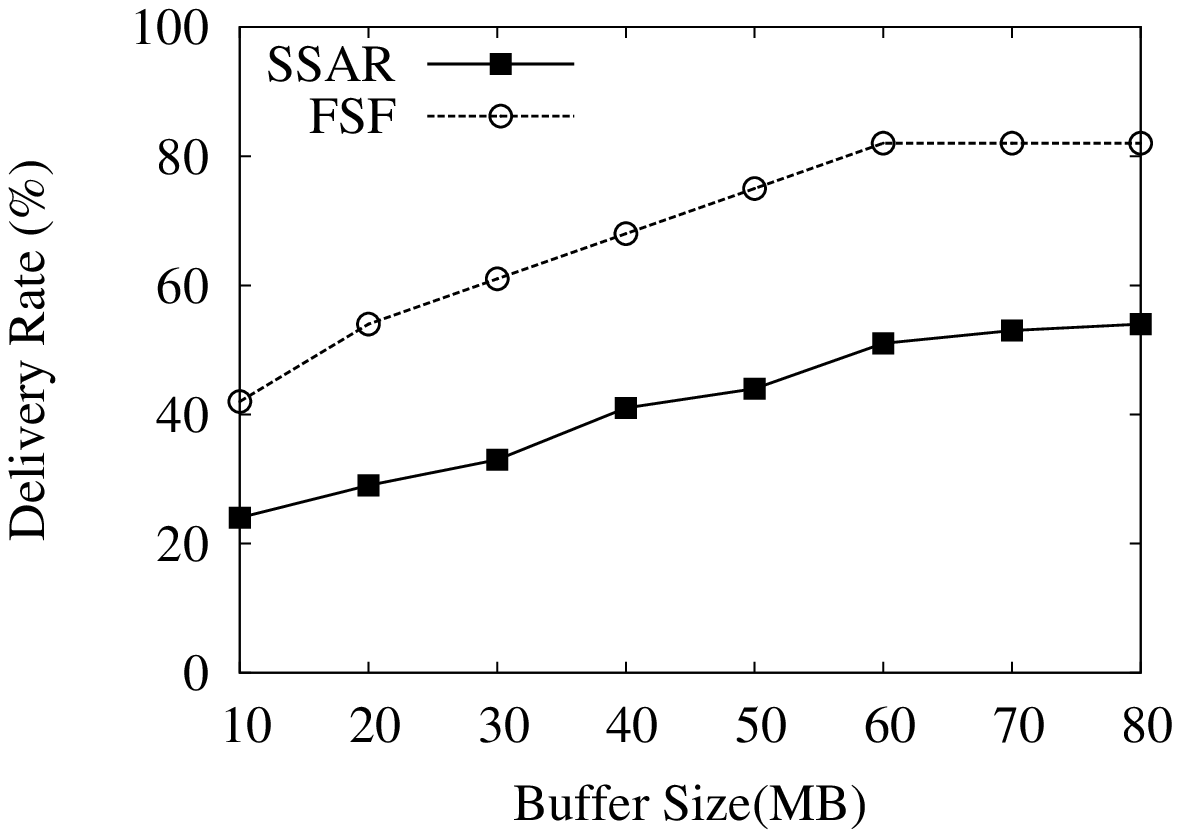}
		\caption{Delivery ratio}
	\end{subfigure}
	\begin{subfigure}[htb]{0.24\textwidth}
		\includegraphics[width=\textwidth]{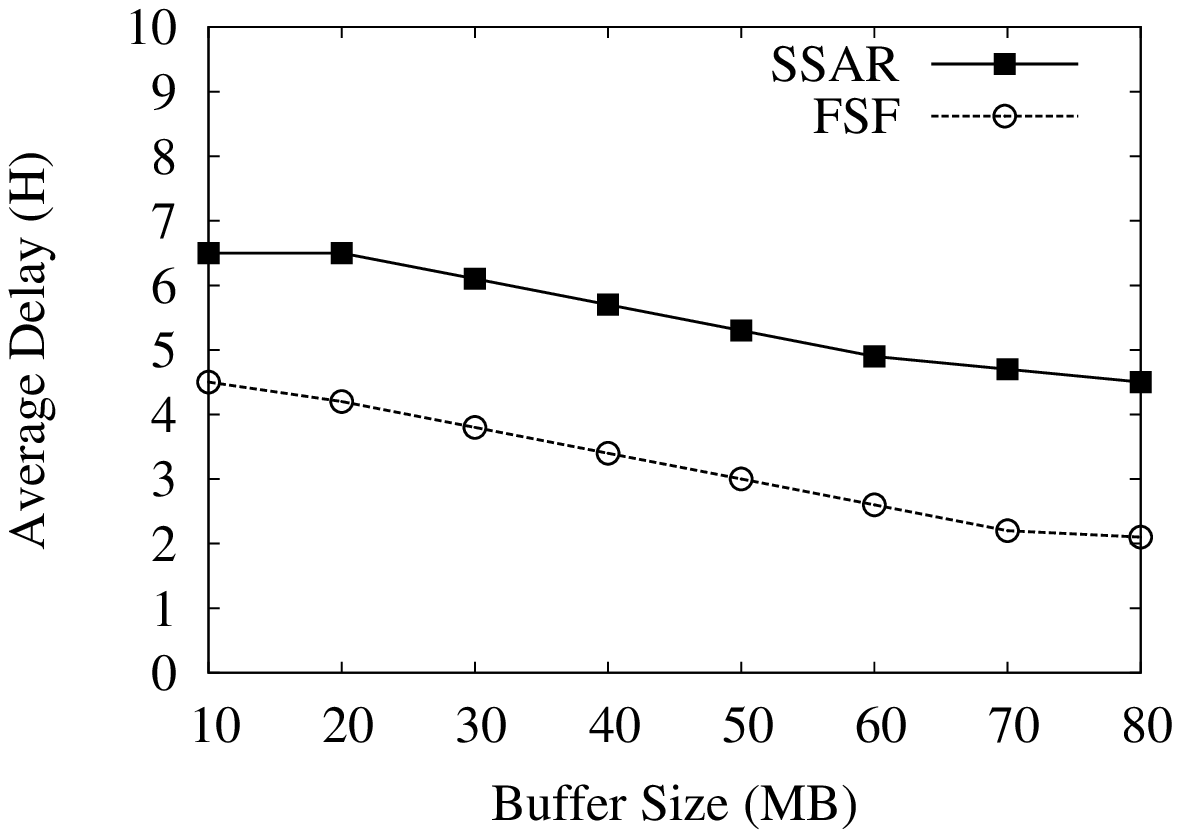}
		\caption{Average Delivery Delay}
	\end{subfigure}
	\begin{subfigure}[htb]{0.24\textwidth}
		\includegraphics[width=\textwidth]{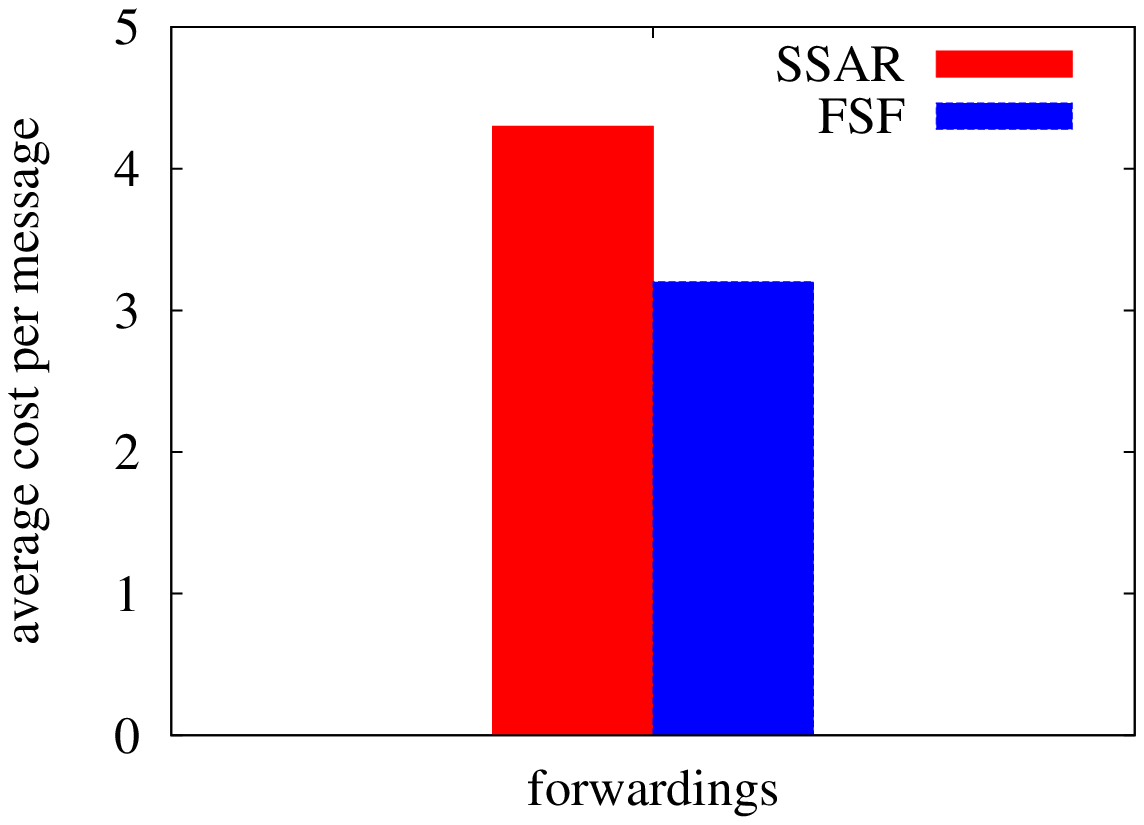}
		\caption{Average Cost}
	\end{subfigure}
	\begin{subfigure}[htb]{0.24\textwidth}
		\includegraphics[width=\textwidth]{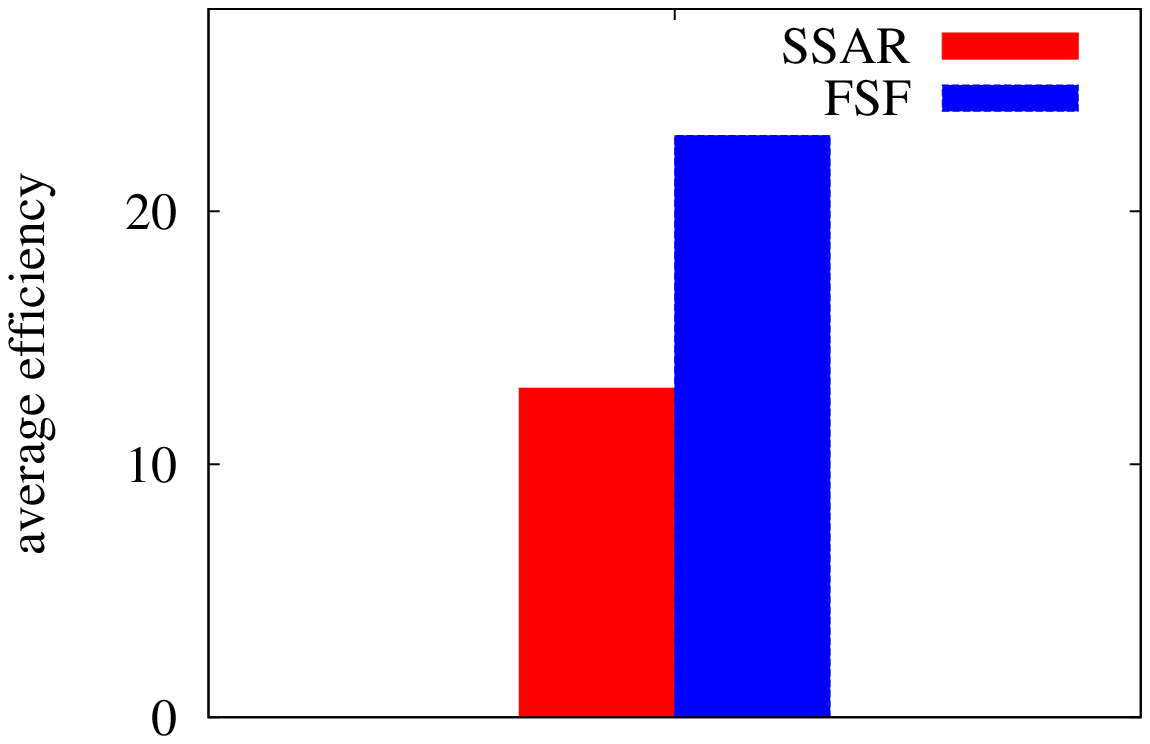}
		\caption{Average Efficiency}
	\end{subfigure}
	\caption{Comparison of FSF x SSAR - simulation results for the Infocom05 scenario.}
	\label{fig:Figure4}
\end{figure*}

Despite the second best result is achieved by the FSF algorithm, it is worth mentioning
that the higher intercontact time among the nodes that have a strong friendship 
can increase the average delivery delay of FSF. The bigger the intercontact among 
these nodes, the higher the average delay, because the relay will need a longer
time to deliver the message to the destination. Thus, FSF can be negatively
impacted if the intercontact time among nodes with a strong friendship is high. 
We are investigating manners to improve FSF performance in such situations by using 
more classes of friendship. Furthermore, another possible solution to this issue is
to take into account the number of encounters among two nodes for routing decisions.
In a contact opportunity, messages will be sent to relays with a strong friendship and with average/higher number of encounters to the message destination.

Figure~\ref{fig:Figure1}.(c) presents the average cost results obtained in the Cambridge scenario. The average cost is the average number of forward events per message during the simulation. The lower the average cost, the better the performance, as all forwarding events consume resources from the nodes involved in the relaying process. Nevertheless, by reducing the number of copies of a message, we reduce the delivery probability. FSF attempts to find a balance between the need for forwarding to improve the delivery ratio, and minimizing transmissions to save resources.  

As we can see from the results in Figure~\ref{fig:Figure1}.(c), FSF achieves the best 
result, lowering the average number of forward events per delivered message compared to
the other evaluated algorithms. It is worth mentioning that the FS algorithm performs 
similarly to FSF. For that purpose, it is reasonable to suppose that considering 
the degree of friendship in routing decisions decreases the average cost per message. These results achieved by both algorithms can be justified by the reduced number of 
messages forwarded when attempting to deliver a message to the destination. The bigger the node willingness to help another node, the higher the delivery probability. In our algorithm, the message will forward to a relay if, and only if, the friendship among the relay and the message destination is strong. As a consequence, the number of message forwarding events will decrease.

Figure~\ref{fig:Figure1}.(d) presents the results for the efficiency metric in the Cambridge scenario. Efficiency is an indicator of network resource usage. The greater the efficiency, the better the utilization of the network resources. Even with a high delivery ratio, a routing algorithm with very low average cost in terms of forwarded messages cannot be considered an efficient algorithm if it consumes excessive network resources.

FSF achieved the best efficiency among all algorithms tested by combining the delivery ratio and the average cost per message. It is worth mentioning that, once again, the FS algorithm achieves the second best result, highlighting the importance of considering friendship strength when taking routing decisions. Despite the higher delivery ratio, the Epidemic algorithm presented less efficiency due to its higher average cost per message.

The FSF effectiveness becomes evident due to the smaller number of forwardings required for delivering a message to each recipient. This occurs because FSF attempts to concentrate on users that have a strong friendship with the message destination, and so it is reasonable to assume that the bigger the friendship strength, the greater the message delivery probability. FSF selects those nodes with a strong friendship with the message destination to take the responsibility for carrying and delivering its messages. In addition, to increase the message delivery probability while saving node resources, less message forwarding events will be required to correctly deliver a message.

\subsection{Results for the Infocom05 scenario}\label{subsec:average_delay}

Figure~\ref{fig:Figure2} presents the results for the Infocom05 scenario. As we can see from this result, the BUBBLE-RAP algorithm achieves the best results in terms of delivery ratio; however, the FSF algorithm performed very similarly to BUBBLE. In addition, FSF achieves this delivery ratio with the half of the average cost per message compared to BUBBLE-RAP. Based on these results, FSF also achieves the best result in terms of average efficiency, as depicted in the Figure~\ref{fig:Figure2}.(d). Considering the results from Figure~\ref{fig:Figure2}, it is reasonable to suppose that, compared to BUBBLE-RAP, FSF improvements are significant. It is worth mentioning that PROPhET and Epidemic achieve the worst performance in the Infocom05 scenario.

Overall, we can justify this outcome by considering that both algorithms, BUBBLE-RAP and FSF, took more advantage of the scenario characteristics. The BUBBLE-RAP algorithm 
combines the knowledge of the community structure with the knowledge of node 
centrality to make forwarding decisions \cite{hui:2011}. Thus, as mentioned before, 
Infocom05 represents the mobility of students participating in the workshops of a 
conference. Thus, based on BUBBLE-RAP steps, it will intuitively create groups of
students by, e.g., country of origin and research topic, and select that most popular
node from each group to carry out messages for the other nodes. Based on the 
characteristics of the Infocom05 trace, the FSF algorithm can classify the relationship
strength among two nodes in the network. For example, based on the aforementioned 
characteristics of Infocom05, it is reasonable to suppose that FSF can 
create two groups of nodes: those nodes belonging to the same country and/or having
the same research topic (have strong friendship), and those belonging to other 
countries or having different research topics (weak friendship).

\subsection{Comparison between FSF and SSAR}\label{subsec:average_cost}

Figures~\ref{fig:Figure3} and \ref{fig:Figure4} present a performance comparison between the FSF and SSAR algorithms. SSAR can work in two modes: forwarding mode, and replication mode. In this work we have used the SSAR algorithm in replication mode because FSF algorithm also functions in that way.

For justifying these results, we need to understand how these algorithms integrate selfishness in their routing decisions, as it may impact network performance. FSF takes into account that messages will not be received in the following cases:

\begin{itemize}
	
	\item  if the friendship strength among the relay and the message destination is weak. In that case FSF considers that people with friendship bonds have more willingness to help each other, and so it attempts to concentrate on these users the responsibility for carrying a message to increase the delivery probability and save node resources.
	
	\item if the relay device resources are at critical levels. It is worth mentioning  that it has a negative impact on the FSF performance because messages addressed to nodes with strong friendship can still be forwarded.  
	
\end{itemize}

In turn, SSAR takes into account that messages will not be received from a relay node in those cases where the relay willingness towards the message destination is not positive (equal to 0). It is worth mentioning that it has also a negative impact on the SSAR
performance, because a user can be interested in carrying messages for very few network users.

Figures~\ref{fig:Figure3} and \ref{fig:Figure4} present the results concerning the delivery ratio in the scenarios evaluated. The FSF algorithm outperforms SSAR in both scenarios evaluated since it takes friendship into account. On the other hand, SSAR takes into account the users' willingness and their contact opportunity to forward messages. Based on the algorithms strategies, it is reasonable to suppose that the FSF strategy potentially favors the delivery of a message, while SSAR meets the users' willingness to carry messages to other nodes, and it can decrease the delivery ratio. We can use this justification with respect to delivery cost and efficiency obtained by both algorithms.

\subsection{Varying Alfa and Beta Thresholds}\label{subsec:efficiency}

\begin{figure*}
	\centering
	\begin{subfigure}[H]{0.24\textwidth}
		\includegraphics[width=\textwidth]{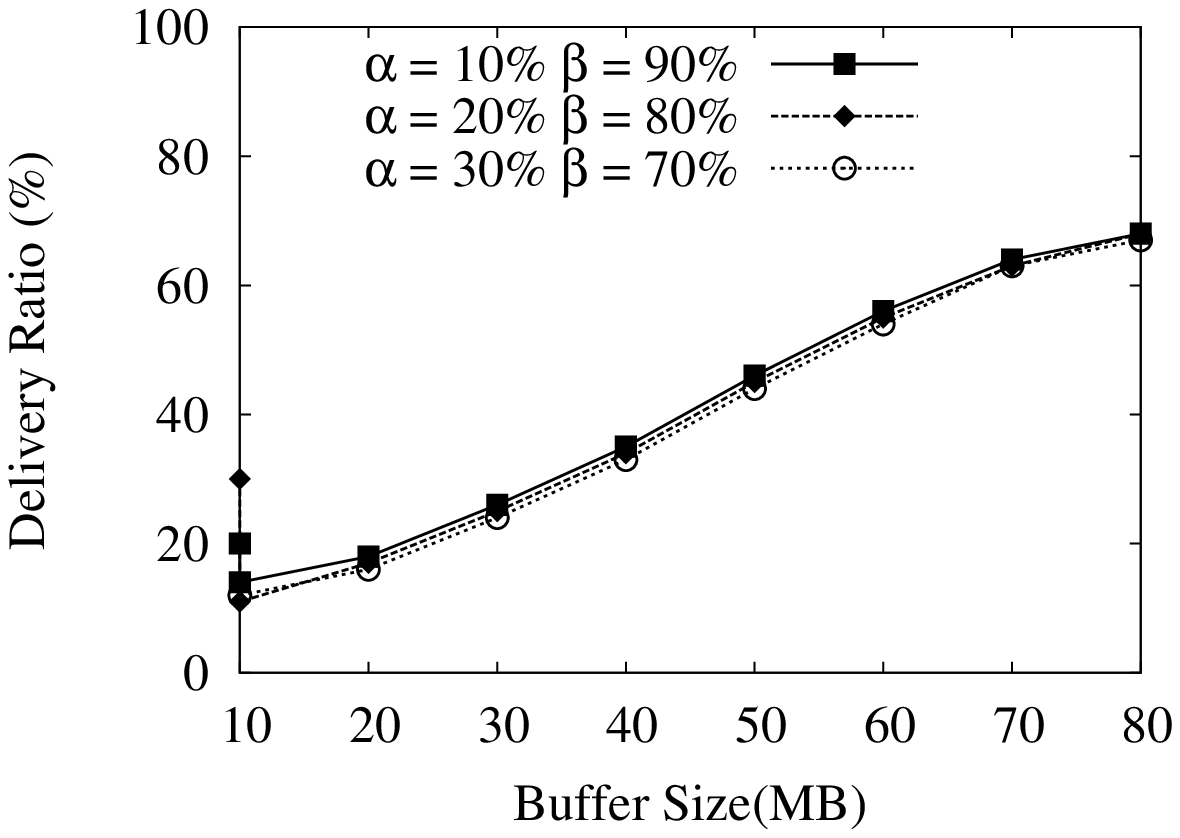}
		\caption{Delivery ratio}
	\end{subfigure}
	\begin{subfigure}[H]{0.24\textwidth}
		\includegraphics[width=\textwidth]{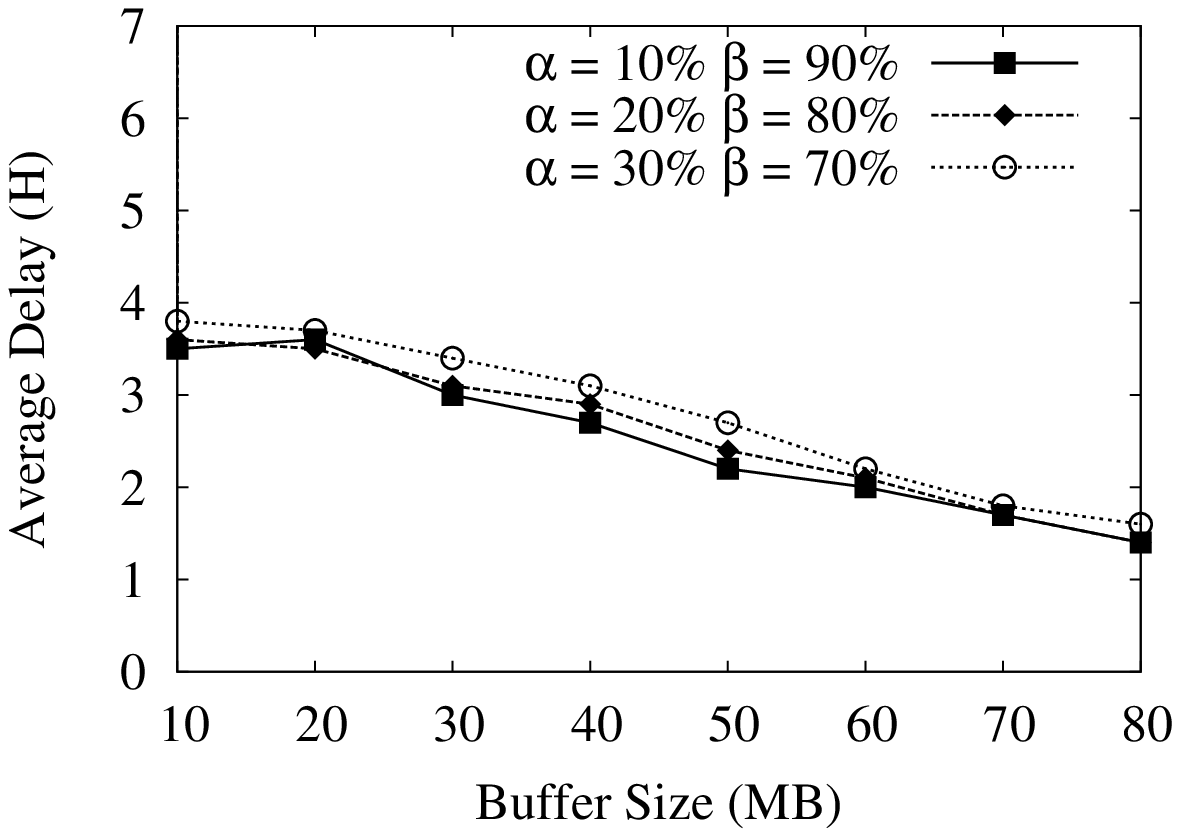}
		\caption{Average Delivery Delay}
	\end{subfigure}
	\begin{subfigure}[H]{0.24\textwidth}
		\includegraphics[width=\textwidth]{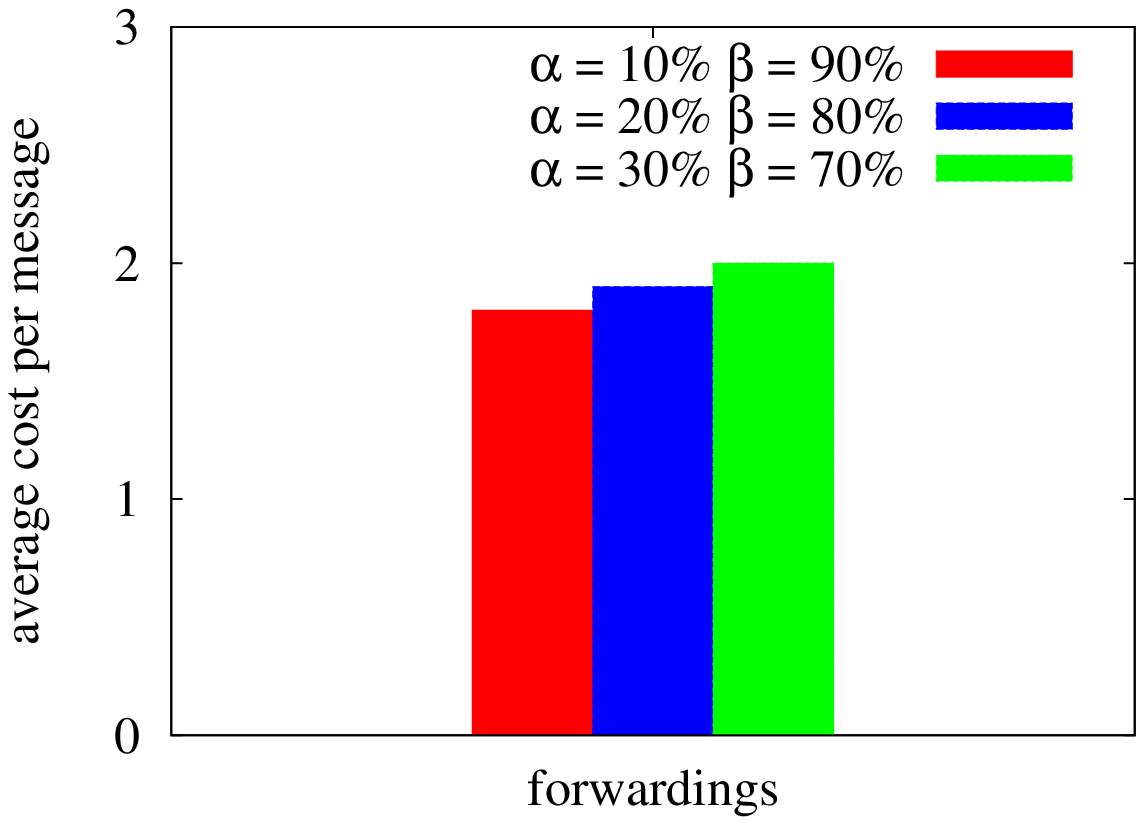}
		\caption{Average Cost}
	\end{subfigure}
	\begin{subfigure}[H]{0.24\textwidth}
		\includegraphics[width=\textwidth]{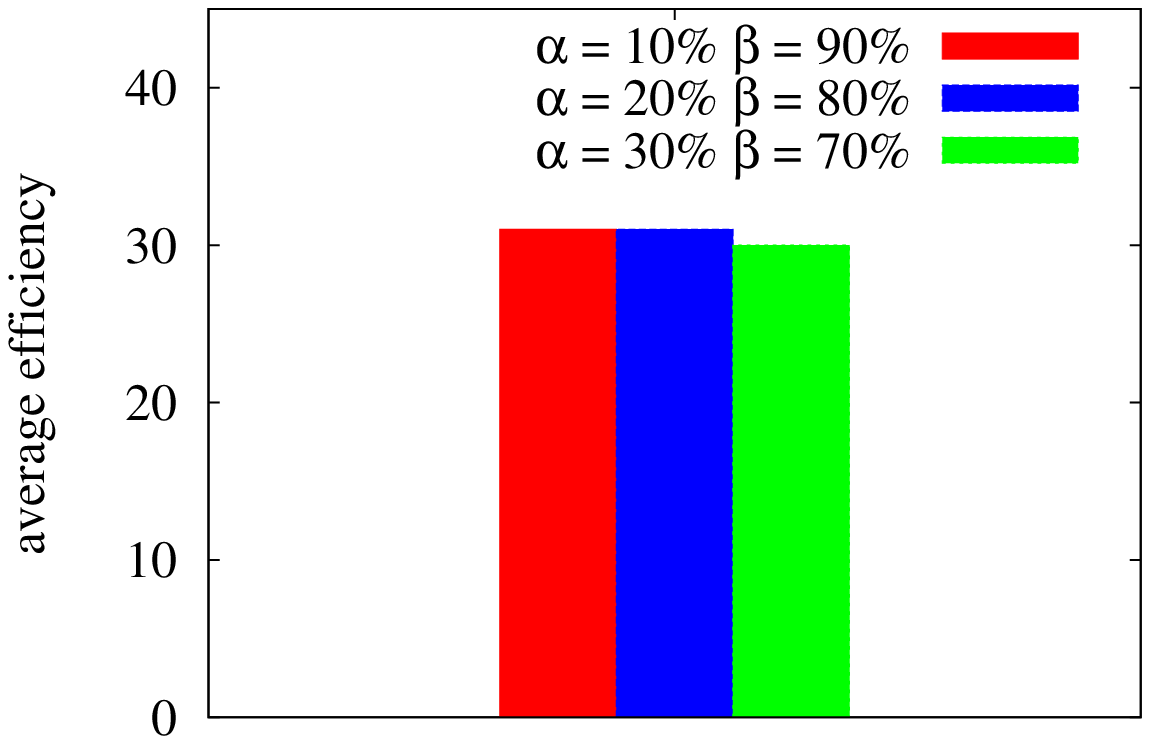}
		\caption{Average Efficiency}
	\end{subfigure}
	\caption{Results for the Cambridge scenario when varying alfa/beta thresholds.}
	\label{fig:Figure5}
\end{figure*}

\begin{figure*}
	\centering
	\begin{subfigure}[htb]{0.24\textwidth}
		\includegraphics[width=\textwidth]{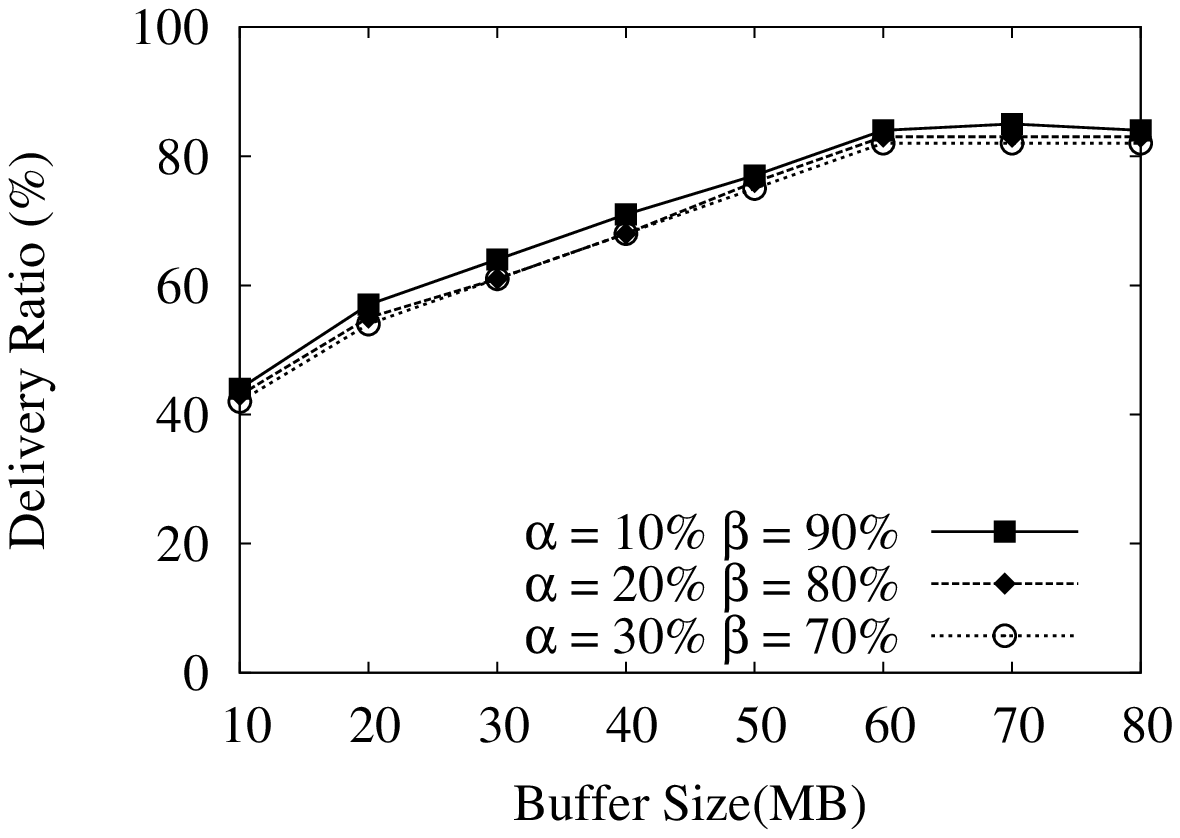}
		\caption{Delivery ratio}
	\end{subfigure}
	\begin{subfigure}[htb]{0.24\textwidth}
		\includegraphics[width=\textwidth]{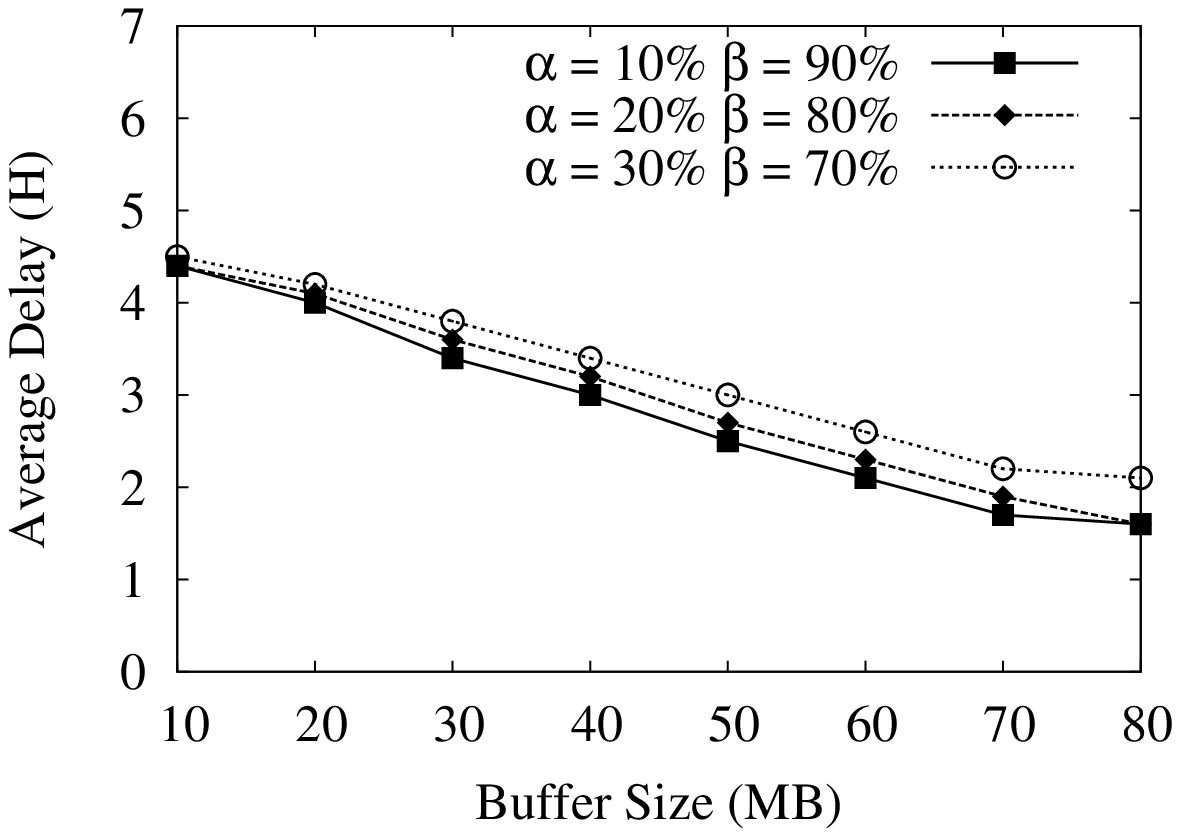}
		\caption{Average Delivery Delay}
	\end{subfigure}
	\begin{subfigure}[htb]{0.24\textwidth}
		\includegraphics[width=\textwidth]{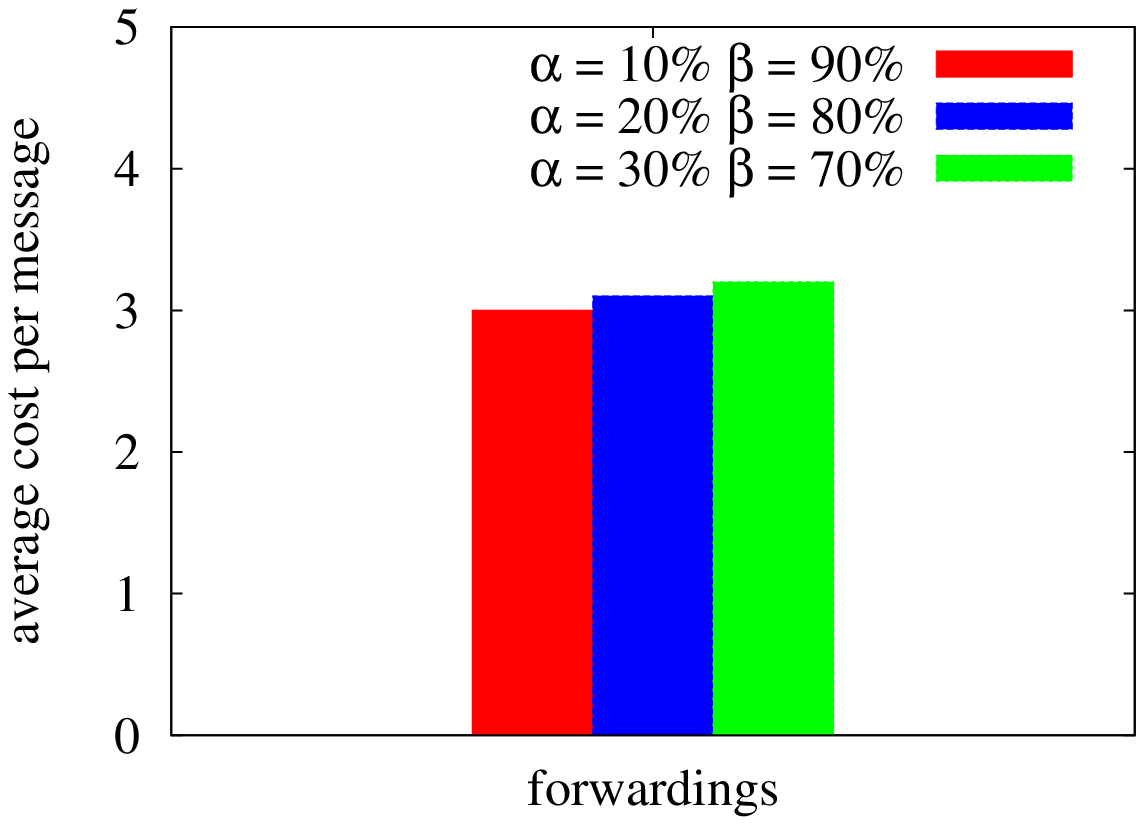}
		\caption{Average Cost}
	\end{subfigure}
	\begin{subfigure}[htb]{0.24\textwidth}
		\includegraphics[width=\textwidth]{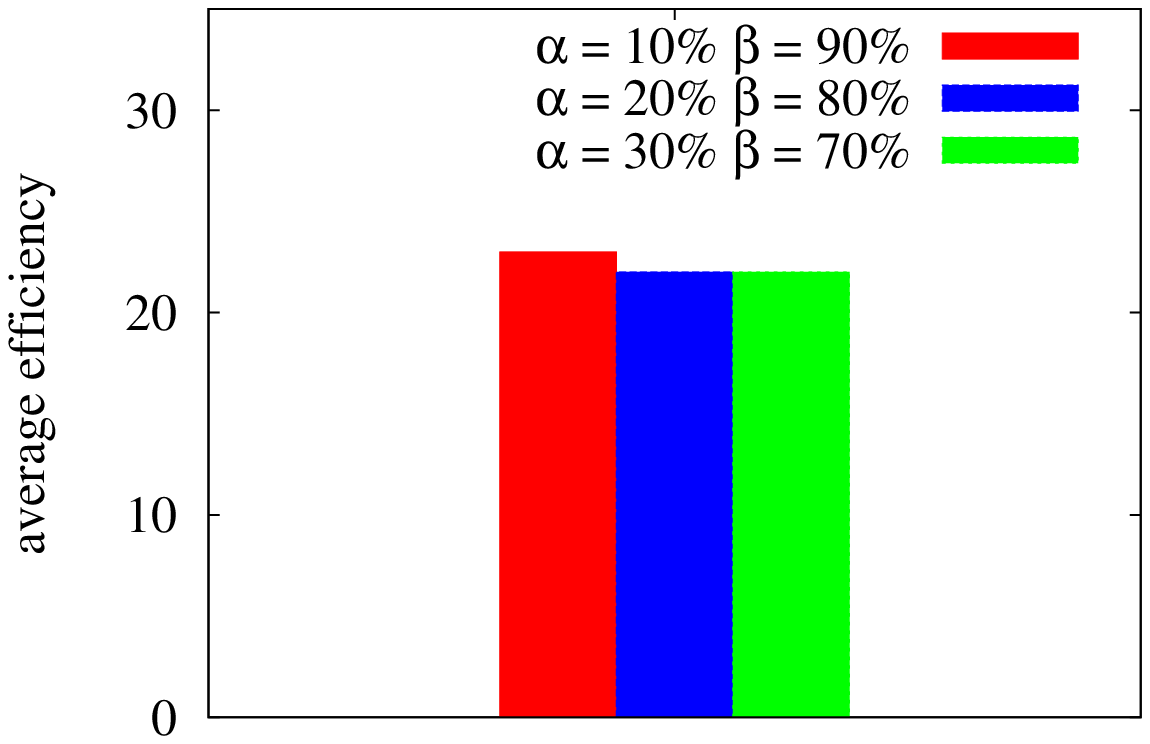}
		\caption{Average Efficiency}
	\end{subfigure}
	\caption{Results for the Infocom05 scenario when varying the alfa/beta thresholds.}
	\label{fig:Figure6}
\end{figure*}

To evaluate different settings for the FSF thresholds, we vary $\alpha\%$ from 10 to 30 and $\beta\%$ from 70 to 90. 
Figures \ref{fig:Figure5} and \ref{fig:Figure6} present the results for the Cambridge 
and Infocom05 scenarios.

We can see that, for all the metrics evaluated, the results are quite
similar. This means that the variation of $\alpha\%$ and $\beta\%$ thresholds 
moderately impacts the FSF performance. Some points of the graph depicting average delivery delay have the more significant differences, e.g. for buffer sizes 30 to 50 in
Figure \ref{fig:Figure5}.(b), and 30 to 80 in Figure \ref{fig:Figure6}.(b). This is caused by the increasing/decreasing of $\alpha\%$ and $\beta\%$ thresholds. On the average 
delivery delay graph, the lower the $\alpha\%$ value and the bigger the $\beta\%$ 
value, the lower is the delivery delay. This means that variations of FSF thresholds
decrease the number of nodes experiencing buffer overflow, and it also decreases the reception probability of messages due to resources constraints. Notice that the bigger the buffer size, the more messages can be stored, and so there are fewer chances of buffer 
overflow. Thus, fewer messages will be dropped and are able to be forwarded in future 
contact opportunities.

\subsection{Additional discussions}\label{subsec:efficiency}

An important issue affecting the performance of the FSF algorithm is the correct 
classification of the friendship among nodes in the network. In this work, we 
introduced an approach to classifying the friendship strength using a machine 
learning algorithm. This algorithm has two important tasks: first, to learn about node
friendship based on data collected from the real world. The second task is to classify new relationships among two nodes in the network. From the FSF results, it is reasonable to assume that Naive Bayes offers an excellent performance at classifying the friendship among nodes. So, the better the classification of the Naive Bayes algorithm, the better the FSF performance.

The deployment of a machine learning algorithm proved to be an interesting solution for the task of classifying friendship strength among two nodes. However, some issues must be taken into account. For example, the availability of information regarding what is friendship in the real world can be a strong requirement since, in some scenarios, this information is not available. To solve this problem, one can monitor the scenario through an application responsible for collecting information about the friendship among nodes. Another alternative is to use a machine learning algorithm that does not need a training database to classify new instances.

Finally, we consider that the use of a machine learning approach can be more flexible than other approaches for classifying the friendship among nodes. For example, if we need to change the friendship model features, using machine learning accelerates this process indeed. Another advantage is that a machine learning approach is able to combine several characteristics to classify the friendship strength, and so it can easily use data from real world situations.

\section{Conclusions and future work}\label{sec:conclusions}

In this work, we proposed the Friendship and Selfishness routing (FSF) algorithm for
Opportunistic Network environments. FSF considers two social characteristics: the 
friendship and the individual selfishness of  candidate nodes for message relaying.
FSF carries out two main tasks: firstly, FSF uses a machine learning technique to 
classify the friendship strength among nodes; secondly, FSF uses a reputation system to assess the relay node selfishness by considering the cases where, despite a strong friendship with the destination, the relay node may decline receiving the message because it is behaving selfishly, or because its device has some resource constraints at that time. 

To validate our proposal, we performed a set of experiments to determine the message delivery effectiveness in two scenarios based on trace-driven simulations using the
ONE simulator.
Compared to two other routing algorithms in the same category, FSF offered better results regarding a set of standard metrics, namely the delivery ratio, the average cost,
and efficiency, while providing similar delivery delay values.

As a future work, we intend to carry out a comparative study by adopting other machine learning algorithms to classify the friendship strength among pairs of nodes.
Moreover, we want to analyze the impact of selfishness in the routing process by using different thresholds for device resource restrictions.


\bibliographystyle{spmpsci}      
\bibliography{lastversion.bib}   

%
%

\end{document}